\documentclass[prc,aps,twocolumn,showpacs,nofootinbib,superscriptaddress,floatfix,preprintnumbers]{revtex4-1}

\usepackage{amsmath,graphicx,mathrsfs,amsfonts}
\usepackage{bm}
\usepackage{subfigure}
\usepackage[colorlinks=true,linktocpage=true,linkcolor=blue,citecolor=blue]{hyperref}
\usepackage[usenames,dvipsnames]{color}

\newcommand{\feq}{f_{\rm eq}}

\newcommand{\taueq}{\tau_r}	

\newcommand{\ped}{{\cal E}}

\newcommand{\pres}{{\cal P}}

\newcommand{\Reducible}{{\cal F}}

\newcommand{\deltaReducible}{(\Reducible_{\rm coll})}

\newcommand{\reducible}{{\mathfrak f}}

\newcommand{\muif}[2]{ { \mu_{#1} \cdots \mu_{#2} } }

\newcommand{\Gen}{{ \Phi }}

\newcommand{\deltaGen}{(\Gen_{\rm coll})}

\newcommand{\gen}{{ \phi }}

\newcommand{\fon}{f_{\rm on}}

\newcommand{\bp}{{\bm{p}}}

\begin{document}

\title{Resummed hydrodynamic expansion for a plasma of particles interacting with fields}

\author{L. Tinti}
\affiliation{Department of Physics, The Ohio State University, Columbus, Ohio 43210, USA}
\affiliation{Institut f\"ur Theoretische Physik, Johann Wolfgang Goethe-Universit\"at, Max-von-Laue-Str.~1, D-60438 Frankfurt am Main, Germany}

\author{G. Vujanovic}
\affiliation{Department of Physics, The Ohio State University, Columbus, Ohio 43210, USA} 

\author{J. Noronha}
\affiliation{Instituto de F\'isica, Universidade de S\~ao Paulo, S\~ao Paulo 05508-090, Brazil} 

\author{U. Heinz}
\affiliation{Department of Physics, The Ohio State University, Columbus, Ohio 43210, USA} 
\affiliation{Theoretical Physics Department, CERN, CH-1211 Gen\`eve 23, Switzerland}
\affiliation{ExtreMe Matter Institute (EMMI), GSI Helmholtzzentrum f\"ur Schwerionenforschung, 
                Planckstrasse 1, D-64291 Darmstadt, Germany}
                
\preprint{CERN-TH-2018-191}

\date{\today}

\begin{abstract}
A novel description of kinetic theory dynamics is proposed in terms of resummed moments that embed information of both hydrodynamic and non-hydrodynamic modes. The resulting expansion can be used to extend hydrodynamics to higher orders in a consistent and numerically efficient way; at lowest order it reduces to an Israel-Stewart-like theory. This formalism is especially suited to investigate the general problem of particles interacting with fields. We tested the accuracy of this approach against the exact solution of the coupled Boltzmann-Vlasov-Maxwell equations for a plasma in an electromagnetic field undergoing Bjorken-like expansion, including extreme cases characterized by large deviations from local equilibrium and large electric fields. We show that this new resummed method maintains the fast convergence of the traditional method of moments. We also find a new condition, unrelated to Knudsen numbers and pressure corrections, that justifies the truncation of the series even in situations far from local thermal equilibrium.
\end{abstract}



\maketitle 

\section{Introduction}
\label{sect:intro}

Relativistic hydrodynamics plays a fundamental role in the description of a wide range of physical phenomena, from astrophysical plasmas to heavy-ion collisions \cite{Bertschinger:1998tv,Mizuno:2015yxa,Heinz:2013th}. There are two main theoretical frameworks for deriving hydrodynamics from an underlying microscopic theory. 

The first one originates from the Chapman-Enskog expansion \cite{ChapmanCowling}, which involves a systematic power counting of the gradients of the standard hydrodynamic quantities, i.e. temperature, chemical potential, and velocity fields. An appealing feature of this approach is that the gradient expansion can be performed even around a quantum field theoretical local equilibrium background, i.e. it does not fundamentally require a classical approximation. Such an expansion can also be done in the relativistic regime, both in the context of kinetic theory \cite{kremer} and in the case of strongly coupled relativistic systems, as shown in Refs.~\cite{Bhattacharyya:2008jc,Baier:2007ix}. 

The straightforward relativistic generalization of the gradient expansion truncated at first order leads to the relativistic Navier-Stokes equations \cite{LandauLifshitzFluids}. These equations violate causality \cite{PichonViscous} and are  linearly unstable around equilibrium \cite{Hiscock_Lindblom_stability_1983,Hiscock_Lindblom_instability_1985} (see also \cite{Denicol:2008ha,Pu:2009fj}). The series can be extended by including second order gradient corrections \cite{Bhattacharyya:2008jc,Baier:2007ix} and also third order terms \cite{GrozdanovKaplisThirdOrder}, though precise statements regarding causality in the nonlinear regime and stability are not available in those cases. Mathematically rigorous results about causality and stability in relativistic viscous hydrodynamics were presented in Ref.~\cite{Bemfica:2017wps} where it was shown that the gradient expansion can be used to derive a causal and stable theory {\it at first order in gradients} if different definitions for the hydrodynamic fields are used. However, recent work showed that the gradient series has zero radius of convergence \cite{Heller:2013fn, Buchel:2016cbj, Denicol:2016bjh, Heller:2016rtz}. Hence, including higher order terms in the expansion does not constitute a viable path towards a systematic improvement of the description of the system in the far-from-equilibrium regime. 

The second method widely used in the derivation of relativistic hydrodynamics employs a moment expansion of the relativistic Boltzmann equation \cite{Denicol:2012cn}. While Boltzmann kinetic theory is only applicable to sufficiently weakly coupled (``dilute'') systems, this approach exploits the idea that hydrodynamics can be understood as an effective field theory describing the long time, long distance macroscopic behavior of the system \cite{Baier:2007ix} whose structure is universal. Therefore, details of the microphysics enter into the hydrodynamic equations only through material properties (such as the equation of state and the transport coefficients) whose calculation involves different methods at weak \cite{Arnold:2000dr} and strong coupling \cite{Son:2007vk}. 

In this approach the Boltzmann equation is expressed in terms of an infinite set of coupled equations for the momentum moments of the the distribution function. Since (at least in the relaxation time approximation, see below) every moment couples only with a finite number of other moments, one can truncate the set of equations at some order, using some approximation for the leftover moments \cite{Denicol:2012cn}. In the absence of long range mean fields, this method can be systematically improved by the inclusion of more moments. Under flow conditions of extreme symmetry where the relativistic Boltzmann equation can be solved exactly \cite{Baym:1984np, Florkowski:2013lza, Florkowski:2013lya, Denicol:2014xca, Denicol:2014tha}, this procedure has been shown to converge rapidly to the exact results of relativistic kinetic theory \cite{Denicol:2014mca, Denicol:2016bjh, Bazow:2016oky}. The reasons behind this rapid convergence even in far-from-equilibrium conditions are, however, still poorly understood as, at first sight, all moments seem to contribute to the equations with similar weight. 

The purpose of this work is explore these questions in greater depth, by extending the method of moments to a more general microscopic background. By generalizing the Boltzmann equation to Boltzmann-Vlasov form we introduce long-range non-collisional forces that could stem from of an electromagnetic gauge field or from medium-dependent particle masses. In the case of electromagnetic interactions the gauge field is calculated self-consistently by taking into account the contribution of the particle current density to Maxwell's equations. We develop the moment expansion of the coupled set of Boltzmann-Vlasov-Maxwell (BVM) equations for a general collision term in 3+1 dimensions, before solving the theory exactly (i.e. with arbitrary numerical precision) in the Relaxation Time Approximation (RTA) \cite{AW} in 0+1 dimensions for a system undergoing boost-invariant longitudinal expansion without transverse flow (Bjorken expansion \cite{Bjorken:1982qr}). 

This paper is organized as follows: In Section~\ref{sect:srdMOM} we briefly review the canonical approach to the method of moments for the Boltzmann equation. In Sec.~\ref{sect:srdBV} we introduce the BVM equations and discuss its expansion in terms of moments. In Sec.~\ref{sect:Res} we propose a new set of moments that resum the contributions of an infinite set of non-hydrodynamic degrees of freedom and show how they can be used to investigate the physics behind the BVM equations efficiently even for massless particle systems. An exactly solvable case of these equations is studied in Sec.~\ref{sect:exactBVM} and compared in Sec.~\ref{sect:comp} with results from the expansion in terms of the resummed moments. We find fast convergence of the resummed moment expansion to the exact solution. Conclusions and a final overview are presented in Sec.~\ref{sect:conclusions}. Various technical discussions are relegated to Appendices A--E.

Unless otherwise stated we use natural units where $\hbar = c = k_B=1$. We adopt the Einstein convention of automatically summing over repeated upper and lower indices, and we represent the contraction (scalar product) between four-vector with a dot: $v^\mu w_\mu = v\cdot w$. The ``mostly minus'' convention for the Minkowski metric is used, i.e. $g^{\mu\nu}={\rm diag}(1,-1,-1,-1)$, as well as the convention $\varepsilon^{0123}=1$ for the four-dimensional Levi-Civita symbol. Round parentheses around groups of Lorentz indices indicate symmetrization, e.g. $A^{(\mu} B^{\nu)} = \frac{1}{2!}(A^\mu B^\nu + A^\nu B^\mu)$.

\section{The method of moments}
\label{sect:srdMOM}

The method of moments, initially introduced in the non-relativistic regime by Grad \cite{grad}, has been widely used to study a number of properties of the relativistic Boltzmann equation \cite{degroot}  
\begin{equation}
\label{RBE}
 p\cdot\partial f = -{\cal C}[f],
\end{equation}
where $f(x,p)$ is the single-particle phase-space distribution function of a gas of particles with mass $m$, and ${\cal C}[f]$ is the collision term which, in general, involves an integral over the distribution function $f$. 

In the moments method one replaces the integro-differential mathematical problem defined by the Boltzmann equation by an  infinite set of coupled partial differential equations for the momentum moments of $f$, which correspond to macroscopic quantities such as, for instance, the fluid's energy-momentum tensor. For the moment we will use Cartesian coordinates such that we do not need to distinguish between the partial derivative $\partial_\mu$ and the covariant one $d_\mu$ \cite{Weinberg:1972kfs}. For any definition of the four-velocity $u^\mu$, we can split $\partial^\mu$ into the time and spatial derivatives in the comoving frame,
\begin{equation}
\label{eq2}  
 \partial^\mu = u^\mu D + \nabla^\mu,
\end{equation} 
where $D{\,=\,}u{\,\cdot\,}\partial$ is the comoving time derivative, also denoted by a dot ($\dot{A}\equiv DA \equiv u^\mu\partial_\mu A$ for any quantity $A$), and $\nabla^\mu\equiv\Delta^{\mu\nu}\partial_\nu$ (where  $\Delta^{\mu\nu}{\,=\,}g^{\mu\nu}{-}u^\mu u^\nu$ projects on the spatial coordinates in the comoving frame). Using these definitions, the Boltzmann equation (\ref{RBE}) can be rewritten as follows \cite{Denicol:2010xn}
\begin{equation}
\label{fdot}
 (p\cdot u) \dot f = -{\cal C}[f] -p{\,\cdot\,}\nabla f,
\end{equation}
from which exact equations of motion for the (reducible) tensor moments
\begin{equation}
\label{reducible_moments}
 \Reducible^{\muif{1}{s}}_r = \int_p (p\cdot u)^r \,p^{\mu_1}\cdots p^{\mu_s} f,
\end{equation}
with $r$ being an integer and $s$ a non-negative integer, can be derived:
\begin{eqnarray}
\label{Red_ev}
 &&\!\!\!\!
 \dot \Reducible^{\muif{1}{s}}_r +\deltaReducible_r^{\muif{1}{s}} =  \\ \nonumber
 &&\!\!\!\!
 = r \dot u_\alpha \Reducible_{r-1}^{\alpha\muif{1}{s}} -\nabla_\alpha\Reducible_{r-1}^{\alpha\muif{1}{s}} 
   + (r{-}1)\nabla_\alpha u_\beta \Reducible_{r-2}^{\alpha\beta\muif{1}{s}}.
\end{eqnarray}
The contribution from the collisional kernel is given by
\begin{eqnarray}
\label{red_coll}
\deltaReducible_r^{\muif{1}{s}} = \int_p (p\cdot u)^{r{-}1}\, p^{\mu_1}\cdots p^{\mu_s}  \, {\cal C} [f].
\end{eqnarray}
To obtain Eq.~(\ref{Red_ev}) one only needs uniform convergence of the momentum integrals (\ref{reducible_moments}).  In (\ref{reducible_moments}), (\ref{red_coll}), and similar integrals below, $\int_p$ indicates the Lorentz invariant momentum integral
\begin{equation}
\label{Lorentz_covariant_integral}
 \!\!\!\!\!
 \int_p \equiv \frac{g}{(2\pi)^3} \int d^4p \, 2\Theta(p_0)\,\delta(p^2{-}m^2) 
           = \frac{g}{(2\pi)^3} \int \frac{d^3p}{p^0},\ 
\end{equation}
with $g$ counting the degeneracy of each momentum eigenstate. In the last expression the integrand must be taken on-shell, i.e. $p^0=\sqrt{m^2{+}\bm{p}^2}$. The particle number current $N^\mu=\int_p p^\mu f$ and energy-momentum tensor $T^{\mu\nu}=\int_p p^\mu p^\nu f$ are given by the moments $ {\cal F}_0^\mu$ and ${\cal F}_0^{\mu\nu}$, respectively. For $(r,s)=(0,2)$, Eq.~(\ref{Red_ev}) gives the following exact evolution equation for the stress-energy tensor: 
\begin{eqnarray}
\label{Tmunu_ev}
   \dot T^{\mu\nu} +\! \int_p \frac{p^\mu p^\nu}{(p\cdot u)}\, {\cal C}[f] 
  = -\nabla_\alpha{\cal F}_{-1}^{\alpha\mu\nu} 
  -\nabla_\alpha u_\beta {\cal F}_{-2}^{\alpha\beta\mu\nu}.\ 
\end{eqnarray}
Its projection onto the four-velocity $u_\mu$ yields four equations describing energy-momentum conservation, $\partial_\mu T^{\mu\nu}{\,=\,}0$ (see Appendix~\ref{sect:divergence}). The remaining six equations provide the exact evolution of the dissipative corrections to the pressure (bulk viscous pressure and shear stress). Unlike the four-momentum conservation equations, the latter couple the components of the stress-energy tensor with other moments of the distribution function.

Since in the comoving frame only the spatial projections of the $T^{\mu\nu}$ couple directly to non-hydrodynamic moments, it is convenient to introduce the following notation for the spatial components of the tensor moment in the local rest frame (LRF) where $u^\mu = (1,\bm{0})$:
\begin{equation}
\label{reduced_moments}
   \reducible^{\muif{1}{s}}_r \equiv \Reducible^{\langle\mu_1\rangle\cdots\langle\mu_s\rangle}_r 
   =\int_p (p\cdot u)^r \,p^{\langle\mu_1\rangle}\cdots p^{\langle\mu_s\rangle} f.
\end{equation}
Angular brackets around a tensor index indicate its spatial components in the LRF obtained by contracting the four-index with the spatial projector $\Delta^\mu_\nu$: $p^{\langle\mu\rangle}\equiv\Delta^\mu_\nu\, p^\nu$. We note that the tensor moments $\Reducible$ are not mutually independent. In fact, the projection with $u_\alpha$ of a tensor moment of rank $(r,s{+}1)$ produces a tensor moment of different rank $(r{+}1,s)$, $u_\alpha \Reducible^{\alpha\muif{1}{s}}_r{\,=\,}\Reducible^{\muif{1}{s}}_{r{+}1}$, while projecting all upper indices along the four-velocity yields a scalar moment $u_{\mu_1} \cdots u_{\mu_s}  \Reducible^{\muif{1}{s}}_r  = \Reducible_{r{+}s} = \reducible_{r{+}s}$. On the other hand, no analogous relation exists for their spatial components in the LRF, i.e. for $\reducible^{\muif{1}{s}}_r$.

It is useful to rewrite the exact evolution equations (\ref{Red_ev}) in terms of $\reducible^{\muif{1}{s}}_r$. As shown in Appendix~\ref{sect:reduction}, their exact evolution equations are 
\begin{eqnarray}
\label{red_ev}
    &&\!\!\!\!\!
      \dot \reducible^{\langle\mu_1\rangle\cdots\langle\mu_s\rangle}_r 
       + \deltaReducible_r^{\langle\mu_1\rangle\cdots\langle\mu_s\rangle}  
       = - \theta \, \reducible^{\muif{1}{s}}_r 
\nonumber\\
    && \!\!
     +\, r \dot u_\alpha \reducible_{r-1}^{\alpha\muif{1}{s}} - s \dot u^{(\mu_1}\reducible_{r+1}^{\muif{2}{s})} 
     - \nabla_\alpha \reducible^{\alpha\langle\mu_1\rangle \cdots \langle\mu_s\rangle}_{r-1}
\nonumber\\
    && \!
      -\, s \nabla_\alpha u^{(\mu_1}\reducible_{r}^{\muif{2}{s})\alpha}  
      + (r{-}1)\nabla_\alpha u_\beta \, \reducible_{r-2}^{\alpha\beta\muif{1}{s}}
\end{eqnarray}
where $\theta\equiv\nabla_\mu u^\mu$ is the scalar expansion rate. These moment equations contain the information we need in this work. For example, by projecting Eq.~(\ref{Tmunu_ev}) with $u_\mu u_\nu$ we obtain Eq.~(\ref{red_ev}) with $r=2$ and $s=0$:
\begin{eqnarray}
\label{eq12}
   \dot\reducible_2 &=& 
   2 \dot u_\mu \reducible_1^\mu - \nabla_\mu \reducible^\mu_1 - \theta \reducible_2 
   + \nabla_\mu u_\nu \reducible^{\mu\nu}_0 
\\ \nonumber 
   &=& \dot u_\mu \reducible_1^\mu - \partial_\mu \reducible^\mu_1 - \theta \reducible_2 
   + \nabla_\mu u_\nu \reducible^{\mu\nu}_0,  
\end{eqnarray}
where we used in the first line that $\int_p p^\mu\, {\cal C}{\,=\,}0$ and in the second line that $u_\mu \reducible^\mu_1{\,=\,}0$. This describes the conservation of energy. Projecting (\ref{Tmunu_ev}) with $u_\mu\Delta^\alpha_\nu$ yields the momentum conservation law
\begin{equation}
\label{eq13}
    \dot\reducible^{\langle\alpha\rangle}_1 = \dot u_\mu \reducible_0^{\mu\alpha} 
    - \dot u^\mu \reducible_2 - \nabla_\mu \reducible^{\mu\langle\alpha\rangle} 
    - \theta \reducible^\alpha_1 -\nabla_\mu u^\alpha \reducible^\mu_1,
\end{equation}
where $\nabla_\mu \reducible^{\mu\langle\alpha\rangle}\equiv \Delta^\alpha_\nu \nabla_\mu \reducible^{\mu\nu}$.

Making use of the general hydrodynamic decomposition of the energy-momentum tensor in the so-called Landau frame \cite{LandauLifshitzFluids}
\begin{equation}
\label{eq14}
  T^{\mu\nu} =  \ped u^\mu u^\nu - \bigl(\pres{+}\Pi\bigr) \Delta^{\mu\nu}   + \pi^{\mu\nu},
\end{equation}
Eqs.~(\ref{eq12}) and (\ref{eq13}) can be rewritten in the more familiar form
\begin{equation}
\label{eq15}
   \dot \ped = -\theta \bigl(\ped{+}\pres{+}\Pi \bigr) +\sigma_{\mu\nu}\pi^{\mu\nu} 
                     ,
\end{equation}
\begin{equation}
\label{eq16}
\begin{split}
    \bigl(\ped{+}\pres{+}\Pi \bigr) \dot u ^\alpha  
    & = \nabla^\alpha \bigl( \pres{+}\Pi \bigr) - \Delta^\alpha_\mu\partial_\nu \pi^{\mu\nu} 
          + \pi^{\alpha\nu}\dot u_\nu,
\end{split}
\end{equation}
where $\ped{\,=\,}\reducible_2$ is the energy density seen by a comoving observer, $\mathcal{E} = u_\mu u_\nu T^{\mu\nu}$, and $\sigma_{\mu\nu} = \Delta_{\mu\nu}^{\alpha\beta}\nabla_\alpha u_\beta$ $\bigl($with the traceless and spatial double projector $\Delta_{\mu\nu}^{\alpha\beta} \equiv \frac{1}{2}\left(\Delta_\mu^\alpha \Delta_\nu^\beta{+}\Delta_\nu^\alpha \Delta_\mu^\beta \right) - \frac{1}{3}\Delta_{\mu\nu}\Delta^{\alpha\beta}\bigr)$ is the shear flow tensor. The shear stress tensor $\pi^{\mu\nu}$ is the traceless part of $\reducible^{\mu\nu}_0$ while the trace of the latter enters the isotropic pressure through $\pres{\,+\,}\Pi= - \frac{1}{3}\Delta_{\mu\nu}\reducible^{\mu\nu}_0$ where $\Pi$ is the bulk viscous pressure. The hydrostatic pressure is given by the equilibrium equation of state $\pres=\pres(\ped,n)$ (where $n$ is the conserved charge if applicable). Therefore, $\reducible^{\mu\nu}_0$ contains all nontrivial information about the dissipative pressure corrections $\Pi$ and $\pi^{\mu\nu}$. Their exact evolution is, therefore, described by the equation
\begin{eqnarray}
\label{eq17}
   && \dot \reducible^{\langle\mu\rangle\langle\nu\rangle}_0 
   + \int_p  \frac{p^{\langle\mu\rangle}p^{\langle\nu\rangle}}{p\cdot u}\,{\cal C}[f] 
   = - \dot u^{(\mu}\reducible_{1}^{{\nu})} 
   -\nabla_\alpha \reducible^{\alpha\langle\mu\rangle \langle\nu\rangle}_{-1} 
\nonumber\\
   &&\qquad\quad -\,\theta \, \reducible^{\mu\nu}_0 
        - 2 \, \nabla_\alpha u^{(\mu}\reducible_{0}^{\nu)\alpha} 
        - \nabla_\alpha u_\beta \, \reducible_{-2}^{\alpha\beta\mu\nu}.
\end{eqnarray}
Differently from the four-momentum conservation equations (\ref{eq12}) and (\ref{eq13}), the equation above couples the components of $T^{\mu\nu}$ with moments of the distribution function that are not part of $T^{\mu\nu}$ or the particle current $N^\nu=\int_p p^\mu f$; this is seen in both the collisional kernel contribution on the left-hand side and the direct couplings on the right-hand side that survive even in the free-streaming limit ${\cal C}[f]\to 0$.

The most common approximation to obtain a closed set of equations is to assume that the deviations from local equilibrium are small and that the deformation of the distribution function is dominated by the non-equilibrium corrections in $T^{\mu\nu}$ and $N^\mu$ \cite{MIS-6}. In this way it is possible to obtain approximate expressions for $\reducible_{-1}^{\alpha\mu\nu}$ and $\reducible^{\alpha\beta\mu\nu}_{-2}$ that can be expressed entirely in terms of quantities usually associated with relativistic hydrodynamic behavior. This procedure leads to Israel-Stewart-like theories \cite{MIS-6} of dissipative relativistic hydrodynamics, as discussed in detail in Ref.\ \cite{Denicol:2012cn}.

It is important to note, however, that thanks to Eq.~(\ref{red_ev}) one can improve the hydrodynamic description by considering $\reducible_{-1}^{\alpha\mu\nu}$ and $\reducible^{\alpha\beta\mu\nu}_{-2}$ as dynamical variables in their own right. The dynamical equations for these moments will then couple to moments of different ranks and orders. This provides a way to systematically improve the solution of the moment equations that does not necessarily depend on the hypothesis of small gradients and approximate local equilibrium \cite{Denicol:2012cn}. In Ref.~\cite{Denicol:2016bjh} such an expansion was tested in $0+1$ dimensions\footnote{%
	In fact, in \cite{Denicol:2016bjh} the local equilibrium expectation values were subtracted from
	the $\reducible$ moments. While this is convenient from the numerical perspective explored 
	in \cite{Denicol:2016bjh}, we note that this does not change the character of the expansion.} 
and fast convergence was found to the corresponding exact solution of the RTA Boltzmann equation. The reasons behind this rapid convergence are not yet fully understood, especially in cases without a large degree of symmetry. 

In the following section we point out that the method of moments becomes much more involved when long-range forces are added to the kinetic system via the coupling to mean fields. Nevertheless, rapid convergence can still be recovered by reorganizing the expansion in terms of the resummed moments presented in Sec.~\ref{sect:Res}.

\section{Adding long-range forces through mean fields}
\label{sect:srdBV}

Extending the results of the previous section to gas mixtures consisting of multiple particle species is conceptually straightforward \cite{degroot}. New difficulties are encountered, however, when one introduces interactions with long-range mean fields, even for a single particle species. Such interactions are described by adding to the relativistic Boltzmann equation an additional force term, the Vlasov term, describing the momentum drift of the distribution function due to acceleration or deflection of particles \cite{kremer}. 

The two most common types of such mean field interactions are medium dependent effective particle masses $m(x)$ and gauge fields. Including them gives rise to the relativistic Boltzmann-Vlasov equation:
\begin{equation}
\label{RBVE}
   p\cdot\partial f + m (\partial_\rho m) \, \partial^\rho_p f + q F_{\alpha\beta}p^\beta\partial^\alpha_p f = -{\cal C}[f].
\end{equation}
Here $\partial^\mu_p{\,\equiv\,}\partial/\partial p_\mu$ is the partial derivative with respect to momentum. In this paper we only consider Abelian $U(1)$ gauge fields with a single conserved charge (electromagnetism) though all mathematical considerations should also apply in the case of non-Abelian gauge fields \cite{Heinz:1983nx,Heinz:1984yq,Heinz:1985qe}. For a recent derivation of dissipative relativistic magnetohydrodynamics from the relativistic Boltzmann equation coupled to Maxwell's equations using the method of moments \cite{Denicol:2012cn} we refer the reader to Ref.~\cite{Denicol:2018rbw}. 

Equation~(\ref{RBVE}) assumes that all four components of $p^\mu$ are independent, i.e. the particle momenta are in general off-shell, with the on-shell constraint imposed by the momentum integration measure in Eq.~(\ref{Lorentz_covariant_integral}) when computing physical quantities. It is sometimes more intuitive to work directly with the on-shell distribution, which depends only on the spatial momenta $\bm{p}$, with $p^0$ being replaced by $\sqrt{m^2{+}\bm{p}^2}$. As shown in Appendix~\ref{sect:on-of}, the covariant equation (\ref{RBVE}) above can be rewritten as the following equation for the on-shell distribution function:
\begin{eqnarray}
\label{RBVE_on}
   p\cdot\partial \fon (x,{\bf p}) &+& m (\partial_i m)\,  \partial^i_p  \fon (x,{\bf p}) 
\\\nonumber
   &+& q F_{i\beta}p^\beta\partial^i_p \fon (x,{\bf p})
   = -{\cal C}_{\rm on}[\fon] (x,{\bf p}).
\end{eqnarray}
However, in this approach one loses manifest relativistic covariance, which makes it difficult to easily switch between the global and fluid rest frames. Also, the derivation of the moment equations becomes more complicated. For these reasons we prefer to use Eq.~(\ref{RBVE}) as our starting point.\footnote{%
	In general, a position dependent mass requires the introduction of an additional mean field $B^{\mu\nu}$
	for thermodynamic consistency, subject to certain constraints to ensure global four-momentum
	conservation and with its own equation of motion \cite{Tinti:2016bav,Romatschke:2011qp}. These 
	extra equations only complicate the algebra but do not change conceptually the following considerations 
	about the method of moments. Therefore we will neglect in the rest of the paper all technical details 
	about the proper implementation of a medium dependent mass and focus mostly on gauge field interactions.}
	
Equation~(\ref{RBVE}) must be solved together with the Maxwell equations that determine the electromagnetic field strength tensor generated by the moving electric charges in the gas:
\begin{equation}
\label{Maxwell}
 \begin{split}
  & \partial_\mu F^{\mu\nu} = J^\nu = q \int_p p^\nu f = q \left( \reducible_1 \, u^\nu + \reducible_0^\nu \right),\\
  & \partial_\mu \tilde F^{\mu\nu} \equiv 
     {\textstyle\frac{1}{2}} \partial_\mu \varepsilon^{\mu\nu\rho\sigma} F_{\rho\sigma} =0.
\end{split}
\end{equation}
Clearly, an external electromagnetic field can also be added if desired (for example one generated by the charges of the colliding nuclei in relativistic heavy-ion collisions). 

Following the same steps as in the preceding section one can rewrite Eq.~(\ref{RBVE}) as the following infinite set of coupled evolution equations for the $\reducible$-moments:
\begin{eqnarray}
\label{red_ev_BV}
  &&  \!\!\!\!\!
       \dot \reducible^{\langle\mu_1\rangle\cdots\langle\mu_s\rangle}_r 
       + \deltaReducible_r^{\langle\mu_1\rangle\cdots\langle\mu_s\rangle} =
       - q \, s \, E^{(\mu_1}\reducible^{{\muif2s})}_{r} 
\nonumber\\
  && -\,q (r{-}1) E_\alpha \, \reducible_{r-2}^{\alpha {\muif 1s}}
                  -q\, s\, \varepsilon^{\rho\sigma\alpha(\mu_1}\reducible_{r-1}^{ {\muif 2s})\beta} 
                   g_{\alpha\beta}u_\rho B_\sigma \ 
\nonumber\\
  && +\, m\dot m \, (r{-}1) \, \reducible_{r-2}^{{\muif 1s}} + s \; m\nabla^{(\mu_1}m \, \reducible_{r-1}^{ {\muif 2s})}
                     -\theta \, \reducible^{\muif{1}{s}}_r  
\nonumber\\
  && +\, r\, \dot u_\alpha \reducible_{r-1}^{\alpha\muif{1}{s}} 
                  - s\dot u^{(\mu_1}\reducible_{r+1}^{\muif{2}{s})} 
                   -\nabla_\alpha \reducible^{\alpha\langle\mu_1\rangle \cdots \langle\mu_s\rangle}_{r-1}
\nonumber\\
  && -\, s \nabla_\alpha u^{(\mu_1}\reducible_{r}^{\muif{2}{s})\alpha} 
        + (r{-}1)\nabla_\alpha u_\beta \, \reducible_{r-2}^{\alpha\beta\muif{1}{s}}.
\end{eqnarray}
Here we used the following relativistic decomposition of the tensor field strength $F_{\mu\nu}$ into the electric and magnetic fields in the comoving frame:\footnote{%
	This is not the only convention found in the literature. The Levi-Civita symbol 
	$\varepsilon_{\mu\nu\rho\sigma}$ is a tensor density of rank $1$, not a tensor 
	\cite{Weinberg:1972kfs}. The magnetic field defined in this way is not a vector 
	but it transforms with an additional determinant of the Jacobian and 
	$g_{\mu\nu}B^\nu = \det(g) B_\mu$. Some authors multiply (divide for the upper 
	case indices) the Levi-Civita by $\sqrt{-\det(g)}$. In this way the magnetic field 
	behaves like a tensor under orientation-preserving transformations. Although in 
	Cartesian coordinates there is no difference, when considering curvilinear coordinate 
	systems (such as Milne coordinates) this fact must be taken into account.}
\begin{eqnarray}
 && F_{\mu\nu} = E_\mu u_\nu - E_\nu u_\mu + \varepsilon_{\mu\nu\rho\sigma} u^\rho B^\sigma,
\end{eqnarray}
with
\begin{eqnarray}
 && E_\mu = F_{\mu\nu} u^\nu, \quad  B_\mu = -\frac{1}{2}\varepsilon_{\mu\nu\rho\sigma} u^\nu F^{\rho\sigma}.
\end{eqnarray}

Compared with Eq.~(\ref{red_ev}), Eq.~(\ref{red_ev_BV}) has five additional terms (the first five terms on the right-hand side) that describe couplings with the mean electromagnetic and mass fields. At first sight it seems straightforward to repeat the procedure described in the previous section to derive evolution equations for the hydrodynamic moments of the distribution function and to systematically improve them by adding the contribution from the non-hydrodynamic moments to which they couple dynamically. However, this simple extension of the moment expansion to the case of the BVM equations (\ref{RBVE},\ref{Maxwell}) faces additional difficulties, as we explain now. 

In the Boltzmann case (\ref{RBE}) the $\reducible$-moments on the right-hand side of Eq.~(\ref{red_ev}) have the same physical dimensions as the ones on the left-hand side: except for the contribution from the collisional kernel, all terms on both sides of the equation have the same sum $r{+}s$ of the energy and tensor rank indices $r$ and $s$. In the Boltzmann-Vlasov case, on the other hand, neither the effective mass $m$ nor the electromagnetic field $F_{\mu\nu}$ are dimensionless; therefore, they couple to $\reducible$-moments of different physical dimensions. Specifically, the first five terms on the right-hand side of Eq.~(\ref{red_ev_BV}) involve $\reducible$-moments whose index sum $r+s$ is lower than that of the moment on the left-hand side. Either the energy index $r$ or the tensorial rank $s$, or both, are reduced. Therefore the systematic improvement of the solution requires considering the influence of moments with ever-decreasing physical dimensions. We note that this already appears in the case of the Boltzmann equation for a single particle species using irreducible moments as shown in Ref.\  \cite{Denicol:2012cn}, where particles with constant mass $m$ were considered. In this case, when $m$ is nonzero there are terms that couple irreducible moments with others of reduced physical dimensions. However, in that case this coupling to moments of lower physical dimensions vanishes for $m=0$. 

In the case of BVM, even for massless particles, this coupling does not vanish because of the electromagnetic interactions.\footnote{%
	This can also be seen from the equations for the irreducible tensor moments first derived in 
	\cite{Denicol:2018rbw}.} 
In the massless limit these lower-dimensional moments become ill-defined for $r{+}s{\,<\,}-2$, and even for nonzero but small masses $m{\,\ll\,}T$ their magnitude grows with increasingly negative values of $r{+}s$, which should affect the convergence of the moment expansion. This is easily seen from their definition (see Eqs.~(\ref{reducible_moments}) and (\ref{reduced_moments})) when writing out the momentum integral in LRF components,
\begin{eqnarray}
   &&
   \reducible_r^{\muif 1s} = 
   \int d^3 p \left(\sqrt{m^2{+}\bp^2}\right)^{r-1} p^{\langle\mu_1\rangle}\cdots  p^{\langle\mu_s\rangle} f 
\\\nonumber
   &&
   = \int_0^\infty dp \; \left(\sqrt{m^2{+}p^2}\right)^{r-1} p^{s+2} 
   \int d\Omega\, \hat{p}^{\langle\mu_1\rangle}\cdots  \hat{p}^{\langle\mu_s\rangle} f,
\end{eqnarray}
where the distribution function $f$ is evaluated on-shell, $p=|\bp|$, $p^0=\sqrt{m^2{+}p^2}$, and $\hat{p}^{\langle\mu_1\rangle}\equiv p^{\langle\mu_1\rangle}/|\bp|$ is a spatial unit vector in the LRF which depends only on the momentum angles $(\theta_p,\phi_p)$, with $d\Omega = \sin\theta_p d\theta_p d\phi_p$. Scaling out the particle mass, $\bp=m\bm{y}$, this becomes
\begin{equation}
\label{suggests}
   \frac{\reducible_r^{\muif 1s}}{T^{r+s+2}} = \left( \frac{m}{T} \right)^{r+s+2}
   \int_0^\infty dy \left(\sqrt{1{+} y^2}\right)^{r-1} y^{s+2}\, \omega^{\muif 1s},
\end{equation}
where the angular integral $\omega^{\muif 1s}\equiv \int d\Omega\, \hat{p}^{\langle\mu_1\rangle}\cdots  \hat{p}^{\langle\mu_s\rangle} f$ is some finite dimensionless number. Eq.~(\ref{suggests}) shows that in the ultra-relativistic limit $m/T{\,\to\,}0$ the $\reducible$-moments diverge for $r{+}s<-2$; for the equilibrium distribution $\feq$ this is easily verified explicitly. In this limit, a systematic solution of the BVM equations requires taking into account the contributions from an infinite set of moments defined by the tower of moment equations in Eq.\,(\ref{red_ev_BV}) as soon as the particle momenta are influenced by non-vanishing mean fields. 

In principle, this problem can be addressed by solving Eq.~(\ref{red_ev_BV}) only for the moments with positive energy index, approximating the other (non-hydrodynamic) moments to which they couple by non-dynamic constitutive equations. This approach was pursued in Ref.~\cite{Denicol:2012cn} where irreducible tensor moments of negative $r$ were expanded in terms of the corresponding irreducible tensors with positive $r$ which are assumed to form a complete basis for the non-equilibrium correction to the distribution function (see  Appendix~\ref{sect:irreducible} for an illustration).

We here propose a different method to avoid couplings to possibly ill-defined moments with $r{+}s{\,<\,}{-}2$, by introducing a new set of moments of the distribution function that resum an infinite number of $\reducible$-moments. We show how all the physical information about the system (including both hydrodynamic and non-hydrodynamic moments of the distribution function) can be recovered from these resummed moments, and that their exact dynamical evolution is well-defined even in the ultra-relativistic $m/T\ll1$ limit. Furthermore, we show for a simplified physical situation that a solution in terms of an expansion in these resummed moments converges rapidly to the exact result from the BVM equations.

\section{Resummed moments and the hydrodynamic expansion}
\label{sect:Res}

Let us define
\begin{equation}
\label{resummed_Gen}
  \Gen^{\muif 1 s}_r(x,\xi^2) \equiv  
  \int_p \, (p\cdot u)^r \, p^{\mu_1}\cdots p^{\mu_s} \, e^{-\xi^2(p\cdot u)^2} f(x,p)
\end{equation}
and introduce the space-like projections
\begin{equation}
\label{resummed_gen}
   \gen^{\muif 1 s}_r = \Gen^{\langle\mu_1\rangle\cdots\langle \mu_s\rangle}_r.
\end{equation}
The similarity of these definitions with Eqs.\ (\ref{reducible_moments}) and~(\ref{reduced_moments}) is obvious: they differ only by the Gaussian weight factor $e^{-\xi^2(p\cdot u)^2}$ under the integral.\footnote{%
	The underlying idea is that this weight function ``generates'' inverse powers of $p\cdot u$ via
	$$\int_0^\infty d\xi\, e^{-\xi^2(p\cdot u)^2} = \frac{\sqrt{\pi}/2}{p\cdot u}.$$}
By Taylor expanding this Gaussian\footnote{%
	There is more than one way to perform such an expansion. One can consider, for instance, 
	$-\xi^2(p\cdot u)^2=x$ and use the Taylor expansion of $e^x$; or $\xi^2(p\cdot u)^2= x^2$ and 
	use the Taylor expansion of the Gaussian $e^{-x^2}$. All such series correspond to an infinite 
	sum of $\reducible_r^{\muif 1 s}$. This is why we call the $\Gen$- and $\gen$-moments 
	``resummed moments'' of the distribution function.}
one sees that each $\gen$-moment $\gen^{\muif 1s}_r$ can be written as an infinite sum of $\reducible$-moments, with the same tensor rank $s$ but different energy indices $r$. The dimensionful parameter $\xi$ determines the relative weight of $\reducible$-moments with different dimensions, i.e. with different factors of $p\cdot u$, in the sum.  

By construction, all resummed moments (\ref{resummed_gen}) of the same tensor rank $s$ are related.\footnote{%
	The Gaussian weight $e^{-\xi^2(p\cdot u)^2}$ falls off quickly enough to preserve uniform 
	convergence in all the following manipulations.}
Using the relations
\begin{eqnarray}
   (p\cdot u)^2 \,  e^{-\xi^2(p\cdot u)^2} & = & -\partial_{\xi^2} \Bigl(e^{-\xi^2(p\cdot u)^2}\Bigr), 
\\ \nonumber 
\\ \nonumber 
  \frac{\sqrt{\pi}}{(p\cdot u)}  e^{-\xi^2(p\cdot u)^2} & = &
  \int_{-\infty}^\infty d\zeta \; e^{-\left(\xi^2 + \zeta^2 \right)(p\cdot u)^2} \nonumber\\
       &=& 2\int_0^\infty d\zeta  \;e^{-\left(\xi^2+\zeta^2 \right)(p\cdot u)^2},
\end{eqnarray}
one easily finds that
\begin{eqnarray}
\label{exact_relations}
 \gen_{r+2}^{\muif 1s}(x,\xi^2) &=& -\partial_{\xi^2} \left[\gen^{\muif 1s}_r(x,\xi^2)\right], \nonumber
\\ 
   \gen^{\muif 1s}_{r-1}(x,\xi^2) &=& \frac{2}{\sqrt{\pi}} \int_0^\infty d\zeta\,\gen^{\muif 1s}_r(x,\xi^2{+}\zeta^2)
\nonumber\\
   &=& \frac{1}{\sqrt{\pi}}\int_{\xi^2}^\infty \frac{d\upsilon}{\sqrt{\upsilon{-}\xi^2}} \; \gen^{\muif 1s}_r(x,\upsilon), \nonumber
\\
   \gen^{\muif 1s}_{r-2}(x,\xi^2) &=& \int_{\xi^2}^\infty d\upsilon \; \gen^{\muif 1s}_r(x,\upsilon), \nonumber
\\
   \gen^{\muif 1s}_{r+1}(x,\xi^2) &=& - \frac{1}{\sqrt{\pi}}\int_{\xi^2}^\infty \frac{d\upsilon}{\sqrt{\upsilon{-}\xi^2}} 
   \, \partial_\upsilon \gen^{\muif 1s}_r(x,\upsilon). \qquad
\end{eqnarray}
The $\reducible$-moments, in particular the hydrodynamic stress-energy tensor and charge current density, are recovered from the resummed moments of the same order and tensor rank via the relation 
\begin{equation}
\label{eq34}
   \reducible_r^{\muif 1s}(x) =\gen^{\muif 1s}_r(x,0).
\end{equation}

Structurally, the exact evolution equations for the resummed moments are very similar to those for the $\reducible$-moments in Eq.~(\ref{red_ev_BV}):
\begin{eqnarray}
\label{gen_ev_BV}
    &&\!\!\!\!\!
        \dot \gen^{\langle\mu_1\rangle\cdots\langle\mu_s\rangle}_r +    
        \deltaGen_r^{\langle\mu_1\rangle\cdots\langle\mu_s\rangle} =
        - q \, s \, E^{(\mu_1}\gen^{{\muif2s})}_{r} 
\nonumber\\
    && -\,q (r{-}1) E_\alpha \gen_{r-2}^{\alpha {\muif 1s}}
         -q\,s\,\varepsilon^{\rho\sigma\alpha(\mu_1}\gen_{r-1}^{ {\muif 2s} ) \beta}g_{\alpha\beta}u_\rho B_\sigma      
\nonumber \\ 
    && +\, m\dot m \, (r{-}1) \, \gen_{r-2}^{{\muif 1s}} + s \; m\nabla^{(\mu_1}m \, \gen_{r-1}^{ {\muif 2s})} 
          - \theta \, \gen^{\muif{1}{s}}_r
\nonumber \\
    && +\,r\, \dot u_\alpha \gen_{r-1}^{\alpha\muif{1}{s}} - s\dot u^{(\mu_1}\gen_{r+1}^{\muif{2}{s})} 
          -\nabla_\alpha \gen^{\alpha\langle\mu_1\rangle \cdots \langle\mu_s\rangle}_{r-1}
\nonumber \\
    && -\, s \, \nabla_\alpha u^{(\mu_1}\gen_{r}^{\muif{2}{s})\alpha}
         + (r{-}1)\nabla_\alpha u_\beta \, \gen_{r-2}^{\alpha\beta\muif{1}{s}} 
\nonumber \\ 
    && -\,2\xi^2\Bigl[ \dot u_\alpha \gen_{r+1}^{\alpha \muif 1s} 
         + \nabla_\alpha u_\beta \, \gen_r^{\alpha\beta\muif 1s} 
\nonumber\\
     &&\qquad\quad  
         +\, m\dot m \, \gen_r^{\muif 1s}
        - q E_\alpha \, \gen_r^{\alpha\muif 1s}\Bigr]. 
\end{eqnarray}
Added complications are the extra $\xi$-dependent coupling terms in the last two lines, and the fact that the resummed moments depend on an additional continuous parameter $\xi$ that effectively adds an extra dimension to the complexity of the numerical solution. This is the price we have to pay in our approach for avoiding the need in the standard approach \cite{Denicol:2012cn} for expanding moments with negative energy index $r$ in terms of positive energy index moments. However, due to the relations (\ref{exact_relations}) between moments of the same tensor rank with different values of $r$, the moment equations can be written entirely in terms of moments with a fixed $r$, thereby removing the need to determine moments with lower $r$ values that could become ill-defined. Hence, in this approach the only truly independent moments are those with different tensor ranks.

Crucially, the moment equations  (\ref{gen_ev_BV}) can be further simplified by using (\ref{exact_relations}) to express everything through moments with $r=1$ such that the terms proportional to $r{-}1$ vanish exactly:
\begin{widetext}
\begin{eqnarray}
\label{gen_ev_BV_r=1}
   \dot \gen^{\langle\mu_1\rangle\cdots\langle\mu_s\rangle}_1 +
   \deltaGen_1^{\langle\mu_1\rangle\cdots\langle\mu_s\rangle} 
   &=& - q \left[ s \, E^{(\mu_1}\gen^{{\muif2s})}_{1} - 2\xi^2\left(  E_\alpha \, \gen_1^{\alpha\muif 1s}
          +  m\dot m \, \gen_1^{\muif 1s} \right) \right] 
\nonumber  \\
   &+&\!\! s \frac{1}{\sqrt{\pi}}\int_{\xi^2}^\infty \frac{d\upsilon}{\sqrt{\upsilon{-}\xi^2}} 
           \left[  m\nabla^{(\mu_1}m \, \gen_{1}^{ {\muif 2s})}
                   -q\, \varepsilon^{\rho\sigma\alpha(\mu_1}\gen_{1}^{ {\muif 2s} ) 
                   \beta}g_{\alpha\beta}u_\rho B_\sigma\right]  
\nonumber \\
    &+&\!\! \frac{1}{\sqrt{\pi}}\int_{\xi^2}^\infty \!\!\! \frac{d\upsilon}{\sqrt{\upsilon{-}\xi^2}} 
            \left[  \dot u_\alpha \gen_{1}^{\alpha\muif{1}{s}} 
                    + s \, \dot u^{(\mu_1}\partial_\upsilon\gen_{1}^{\muif{2}{s})} 
                    + 2\xi^2 \, \dot u_\alpha \, \partial_\upsilon \gen_{1}^{\alpha \muif 1s}  
                    - \nabla_\alpha \gen^{\alpha\langle\mu_1\rangle \cdots \langle\mu_s\rangle}_{1} \right] 
\nonumber \\
    &-&\!\! \theta \, \gen^{\muif{1}{s}}_1 - s \, \nabla_\alpha u^{(\mu_1}\gen_{1}^{\muif{2}{s})\alpha} 
         - 2\xi^2\nabla_\alpha u_\beta \, \gen_1^{\alpha\beta\muif 1s}.  
\end{eqnarray}
\end{widetext}
Here the resummed moments under the integrals are understood as functions of the auxiliary variable $\sqrt{\upsilon}$ instead of $\xi$. These equations are well-defined for any tensor rank $s$, even in the massless limit. The expansion in terms of resummed moments whose dynamics follows Eq.~(\ref{gen_ev_BV_r=1}) therefore provides a well-defined and systematically improvable generalization of the hydrodynamic expansion of the Boltzmann-Vlasov equation in the presence of mean field forces. 

It is worth noting that after integrating Eqs.~(\ref{gen_ev_BV_r=1}) over $\xi$ one recovers Eq.~(\ref{red_ev_BV}) for the $\reducible$-moments with energy index $r{\,=\,}1$ \textit{exactly}. More generally, with appropriate manipulations using relations (\ref{exact_relations}) one recovers the evolution equations for all the well-defined $\reducible$-moments, before any approximations (in particular the lowest-order ones that form the equations of hydrodynamics in the presence of mean fields).\footnote{%
	By applying consistent approximations to the higher-order, non-dynamical moments (see 
	discussion below) one can ensure that the transport coefficients are the same in both approaches.}
However, by using the resummed $\gen$-moments instead of the $\reducible$-moments we avoid the numerical issues associated with $\reducible$-moments with sufficiently negative energy index.

For the simultaneous solution of the Maxwell equations (\ref{Maxwell}) one expresses the electric current in terms of the resummed moments as follows (see Eq.~(\ref{eq34})):
\begin{equation}
   J^\nu(x) = q \biggl(u^\nu(x)\, \phi_1(x,0) + \frac{2}{\sqrt{\pi}} \int_0^\infty \!\! d\xi \, \gen_1^\nu(x,\xi) \biggr).
\end{equation}
%

\section{Exactly solvable case: electromagnetic plasma in 0{\,+\,}1 dimensions}
\label{sect:exactBVM}

To test numerically the convergence properties of the set of equations (\ref{gen_ev_BV_r=1}) we study a situation where one can solve the BVM equations exactly. The approximations for physically meaningful macroscopic quantities generated by truncating the moment expansion can then be checked order by order.

We will focus on the massless case where the complications mentioned in the previous section are most relevant.  In the absence of mean-field forces an exact solution of the Boltzmann equation in RTA was found in Refs.~\cite{Baym:1984np,Florkowski:2013lza,Florkowski:2013lya} for a transversally homogeneous gas undergoing longitudinally boost-invariant expansion, i.e. Bjorken flow \cite{Bjorken:1982qr}. For this case approximations made within the traditional moment expansion \cite{Denicol:2012cn} for the dynamics of the shear stress tensor are known to provide a very good description of the exact result obtained from the Boltzmann equation \cite{Alqahtani:2017mhy}, for reasonable values of transport coefficients. We now show that this exact solution can be extended to the RTA Boltzmann-Vlasov case, which will be used as our testing ground.

\subsection{Exact solution of the BVM equations for Bjorken symmetry}
\label{sec5a}

Because of the symmetries of Bjorken flow it is convenient to formulate the problem in Milne coordinates $x^\mu=(\tau, x, y, \eta)$, with metric $g_{\mu\nu}{\,=\,}\mathrm{diag(1,-1,-1,-\tau^2)}$, where
\begin{equation}
  \tau = \sqrt{t^2{-}z^2}, \qquad \eta=\frac{1}{2}\ln\left( \frac{t{+}z}{t{-}z} \right).
\end{equation}
The quantities $\tau$ and $\eta$ are called, respectively, longitudinal proper time and space-time rapidity or, in short, proper time and rapidity. Correspondingly, we use for the four-momentum $p^\mu=(p^\tau, p^1, p^2, p^\eta)=g^{\mu\nu} p_\nu =(p_\tau, -p_x, -p_y, - p_\eta/\tau^2)$. The mass-shell condition for massless particles reads $p_\tau{\,=\,}\sqrt{\bm{p}_T^2{+}p_\eta^2/\tau^2}$ where $-\tau p^\eta=p_\eta/\tau{\,=\,}p_z{\,=\,}p_L$ is the longitudinal momentum component in Cartesian coordinates. The Cartesian momentum-space volume element $d^3p \equiv d^2p_T\, dp_L$ is written in Milne coordinates as $d^2p_T\,dp_\eta/\tau\equiv d^3\tilde{p}/\tau$. For more information about the underlying symmetries of the Bjorken expanding fluid in hydrodynamics and kinetic theory see Refs.\ \cite{Gubser:2010ze} and \cite{Denicol:2014xca,Denicol:2014tha}, respectively.

The symmetry of the expansion imposes that the (on-shell) distribution functions of all particle species depend only on the proper time, the transverse momentum $p_T$, and the longitudinal momentum $p_\eta$. If the electromagnetic field is dynamical and there are no external sources, Bjorken symmetry also constrains the fields and the electric current. The simplest case consistent with all symmetry constraints is an overall charge neutral two-component gas of particles $f(\tau, p_T,p_\eta)$ and antiparticles $\bar f(\tau, p_T,p_\eta)$, expanding in a purely longitudinal electric field $E_\eta(\tau)$ without magnetic components, and evolving according to the following equations for the on-shell distributions and electric field:
\begin{eqnarray}
\label{particle_BV} 
 && \partial_\tau f + q E_\eta \frac{\partial f}{\partial p_\eta} 
       = -\frac{1}{\taueq} \left( \vphantom{\frac{}{}} f{-}\feq\right),  
\\
\label{antiparticle_BV}
 &&  \partial_\tau \bar f - q E_\eta \frac{\partial \bar f}{\partial p_\eta} 
        = -\frac{1}{\taueq} \left( \vphantom{\frac{}{}} \bar f{-}\feq\right),
\\
\label{exact_E_Maxwell}
 && \partial_\tau\left(\frac{E_\eta}{\tau}\right) 
      = - q \int \frac{ d^3\tilde{p}}{\tau} \, \frac{p_\eta}{\tau p^\tau} \left( \vphantom{\frac{}{}} f{-}\bar f \right).  
\end{eqnarray}
Here $\taueq$ is the relaxation time present in the RTA collision term. To preserve the conformal symmetry of this massless plasma we assume $\taueq(\tau)=c/T(\tau)$ where $T(\tau)$ is the effective temperature of the system at proper time $\tau$ and the constant $c$ is related to the specific shear viscosity $\bar\eta\equiv\eta/s$ (i.e. the ratio of shear viscosity to entropy density) of the system by $c=5\bar\eta$ \cite{Denicol:2010xn,Denicol:2011fa,Denicol:2014xca,Denicol:2014tha}. The equilibrium distribution function reads, for both particles and antiparticles,
\begin{equation}
   \feq(\tau,p_T,p_\eta) = \exp\left[ -p\cdot u(\tau)/T(\tau) \right].
\end{equation}
For a system undergoing Bjorken expansion the flow four-velocity reduces to $u^\mu=(1,\bm{0})$ in Milne coordinates. The effective temperature $T$ is defined through the usual Landau matching prescription $\ped=\ped_{\rm eq} = 48\pi k T^4$ where $k\equiv g/(2\pi)^3$ (see Eq.~(\ref{Lorentz_covariant_integral})).
Appendix~\ref{sect:exact_BVM_equations} presents an explicit derivation of the evolution equations and the demonstration that  the symmetries ensure that in Milne coordinates the flow remains static at all proper times.

The system of equations (\ref{particle_BV})-(\ref{exact_E_Maxwell}) is very similar to the one studied in Ref.~\cite{Ryblewski:2013eja}, except that we here consider a $U(1)$ electromagnetic gauge field (which is not related to an Abelian subgroup of color $SU(3)$) and also neglect the effects from nonperturbative electric field decay, which are induced via the Schwinger mechanism. The latter is exponentially suppressed by the inverse electromagnetic coupling $\alpha^{-1}$, and thus in heavy-ion collisions the electromagnetic Schwinger effect acts on a much slower time scale than those characterizing the strong interactions.

The solution of Eqs.~(\ref{particle_BV}) and~(\ref{antiparticle_BV}) for the particle and antiparticle distributions reads
\begin{eqnarray}
\label{part}
   f(\tau,p_T,p_\eta) &=& D(\tau,\tau_0)f_0\bigl(p_T,p_\eta{+}\Delta p_\eta(\tau,\tau_0)\bigr) 
\\\nonumber
   &+& \int_{\tau_0}^\tau ds \frac{D(\tau,s)}{\taueq(s)} \feq\bigl(s,p_T,p_\eta{+}\Delta p_\eta(\tau,s)\bigr), 
\\
\label{antipart}
   \bar f(\tau,p_T,p_\eta) &=& D(\tau,\tau_0)\bar f_0\bigl(p_T,p_\eta{-}\Delta p_\eta(\tau,\tau_0)\bigr)
\\\nonumber
    &+& \int_{\tau_0}^\tau ds \frac{D(\tau,s)}{\taueq(s)} \feq\bigl(s,p_T,p_\eta{-}\Delta p_\eta(\tau,s)\bigr).
\end{eqnarray}
The initial conditions are $f=f_0$ and $\bar f =\bar f_0$ at $\tau = \tau_0$. The damping function $D$ and the momentum shift $\Delta p_\eta$ are defined as
\begin{equation}
 D(\tau_2,\tau_1) = \exp\left[ -\int_{\tau_1}^{\tau_2} ds \, \frac{1}{\taueq(s)} \right],
\end{equation}
\begin{equation}
 \Delta p_\eta (\tau_2,\tau_1) =  q \int_{\tau_1}^{\tau_2} ds\,  E_\eta(s). 
\end{equation}
%
%
\begin{figure*}[t!]%
\begin{center}
\includegraphics[angle=0,width= 0.8\linewidth]{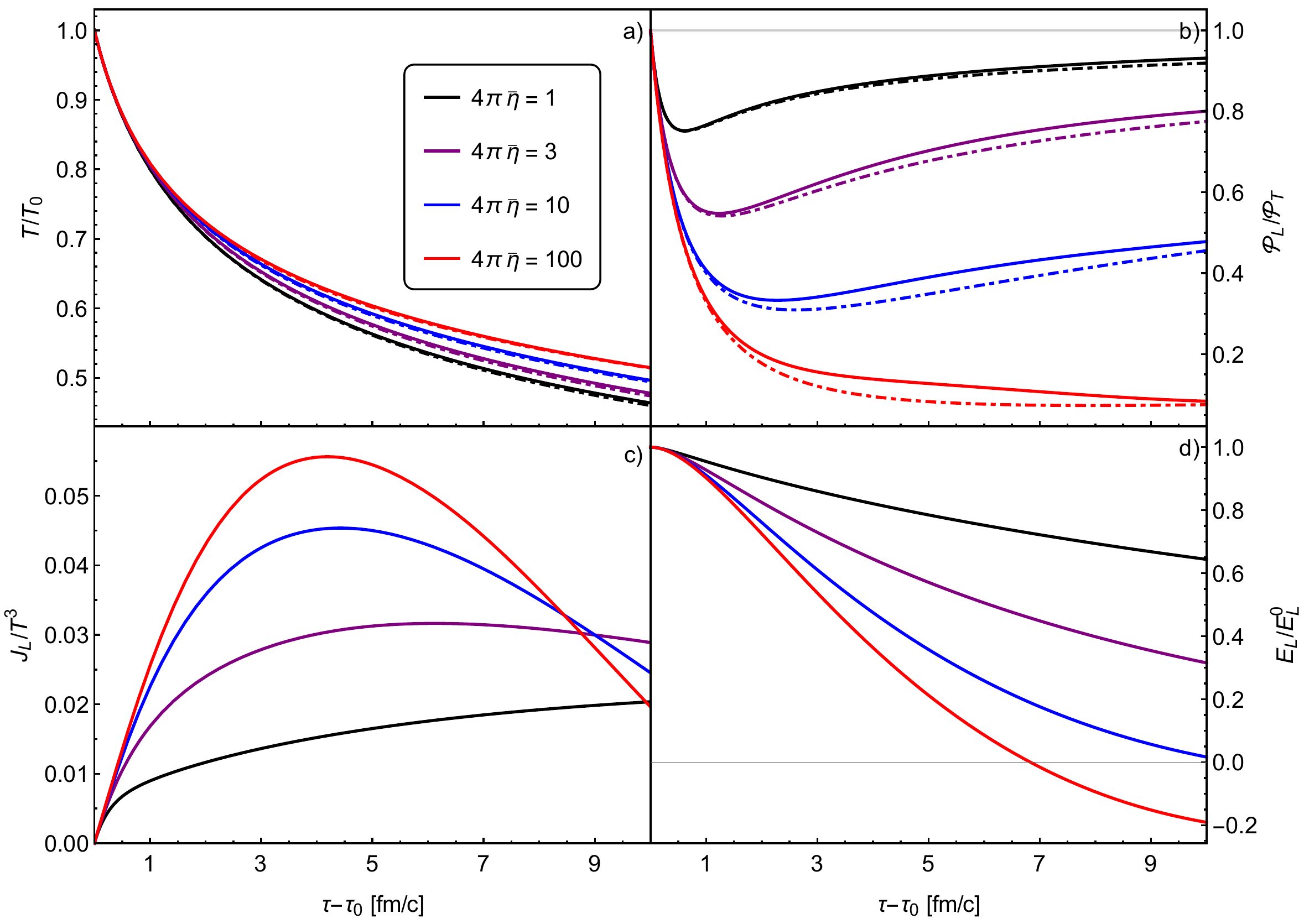}
\end{center}
\caption{	
	(Color online) Time evolution of the temperature (a), pressure anisotropy (b), electric current (c), and
	electric field (d), for four different choices of the specific entropy $\bar\eta=\eta/s$ (denoted by different 
	colors as described in the legend) and two different choices of the initial longitudinal electric field, 
	$E_\eta^0/T_0=0$ (solid lines) and $E_\eta^0/T_0=1/5$ (dash-dotted lines). The initial temperature 
	was chosen as $T_0=0.3$\,GeV at starting time $\tau_0=1$\,fm/$c$.
	}
\label{F1}
\end{figure*}%
%
The implicit equations (\ref{part}) and~(\ref{antipart}) can be solved, for arbitrary initial distributions $f_0$ and $\bar f_0$, by numerically iterating the temperature and electric field profiles, $T(\tau)$ and $E_\eta(\tau)$, subject to the Landau matching constraint
\begin{equation}
\label{exact_temp}
   48 \pi k \, T^4(\tau) = \ped(\tau) = \int \frac{d^3\tilde{p}}{\tau} \, p_\tau \left( \vphantom{\frac{}{}} f{+}\bar f  \right)
\end{equation}
and the solution to the Maxwell equation (\ref{exact_E_Maxwell})
\begin{equation}
\label{exact_el}
   E_\eta (\tau) = \frac{\tau}{\tau_0}E_\eta^0  
   - q \tau \int_0^\tau ds \int \frac{d^3\tilde{p}}{s} \, \frac{p_\eta}{s p^\tau} \left( \vphantom{\frac{}{}} f{-}\bar f \right),
\end{equation}
following the method used in Ref.~\cite{Ryblewski:2013eja}: One inserts an initial guess for the temperature and electric field profiles into the right-hand sides of Eqs.~(\ref{part}) and (\ref{antipart}) and then uses the distribution functions obtained from these equations to obtain new profiles from Eqs.~(\ref{exact_temp}) and (\ref{exact_el}). The process is iterated until convergence of the profiles, at some desired precision, is achieved. The resulting particle and antiparticle distribution functions form (for all practical purposes) an exact solution of the coupled BVM equations for systems with Bjorken symmetry. 

As a basis for comparisons with the resummed expansion discussed further below, we consider a set of thermal equilibrium initial conditions with parameters that yield energy densities that can be considered reasonable for problems of interest in the field, such as high energy proton-antiproton collisions. We take for the initial temperature $T_0 =0.3$\,GeV defined at the initial time $\tau_0 = 1$\,fm/$c$. For the initial electric field we assume a temperature-normalized ratio $E_\eta^0/T_0 = (\tau_0 E_L^0)/T_0=1/5$, corresponding to a normalized value for the longitudinal Cartesian component of the initial electric field given by $E_L^0/T_0^2{\,\approx\,}2/15$. In dimensionful units this initial electric field has the value $E_L^0{\,\approx\,}0.3$\,fm$^{-2}$. This is a natural order of magnitude in this case because it corresponds to the electric field of a parallel plate capacitor made of conducting sheets representing the central collision between a proton and an antiproton that are both (infinitely) Lorentz contracted, with charge densities $\sigma$ corresponding to the charge of a proton smeared over a disk of radius 1\,fm: $E = \sigma = 1/(\pi r_p^2) \approx (1/\pi)$\,fm$^{-2}$. 

Figures~\ref{F1}a-d show the evolution of the temperature $T/T_0$, longitudinal/transverse pressure ratio $\pres_L/\pres_T$,\footnote{%
	The longitudinal pressure is the projection of the stress-energy tensor on the direction 
	$z^\mu$, i.e. $\pres_L = T^{\mu\nu}z_\mu z_\nu$ reduces to $T^{33}$ in the local rest 
	frame where $z^\mu=-\delta^\mu_3$. Since for a conformal system the bulk viscous 
	pressure vanishes, the transverse pressure can then be computed from the 
	energy-momentum tensor of the particles as $\pres_T = - \frac{1}{2}
	\left(\Delta_{\mu\nu}T^{\mu\nu} + \pres_L\right)$.
	}
normalized electric current $J_L/T^3$, and electric field $E_L/E_L^0$, respectively, for four different choices of the specific shear viscosity, $4\pi \bar \eta =1,\,3,\,10,$ and 100, covering the range from the KSS result \cite{Kovtun:2004de} to almost free-streaming. We compare the evolution according to the BVM equations (solid lines) with that of a system of uncharged particles, i.e. without electric field (dash-dotted lines).

%
\begin{figure*}[t]%
\centering
\includegraphics[angle=0,width=0.8\linewidth]{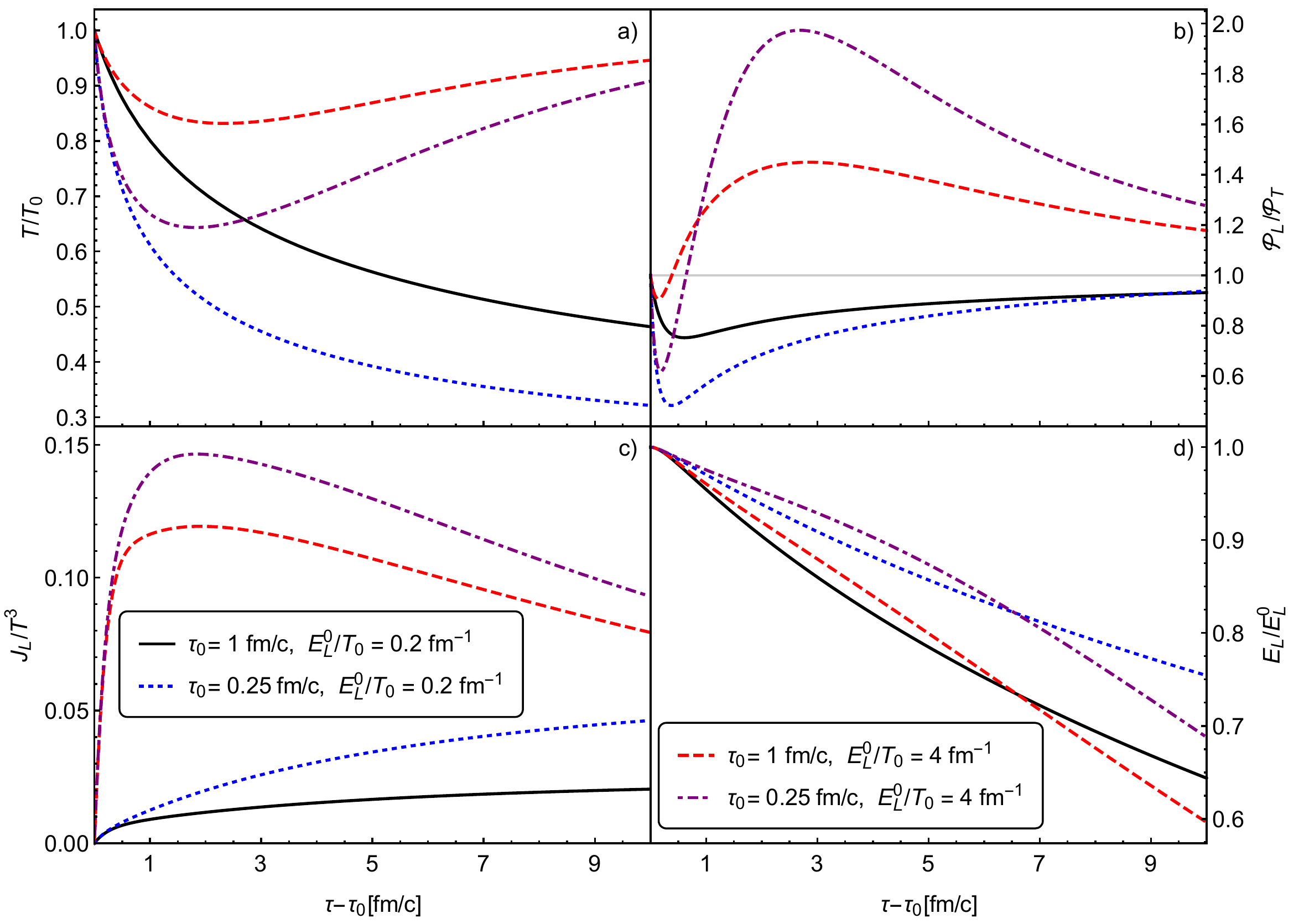}
\caption{
	(Color online) Time evolution of the temperature (a), the electric field (b), the pressure anisotropy  
	$\pres_L/\pres_T$ (c), and the electric current (d), for fixed initial temperature $T_0=0.3$\,GeV 
	and minimal specific shear viscosity $4\pi\bar\eta=1$ but two different choices of the initial electric 
	field that differ by a factor 20 and two different starting times $\tau_0$ that differ by a factor 4 
	(colored lines, see legend in panel (d)). The black solid lines show for comparison the results from 
	Fig.~\ref{F1} for the same shear viscosity.  
	}
\label{F2}
\end{figure*}
%

According to Fig.~\ref{F1}a the temperature evolution is not significantly affected by the presence of the electric field. This is easily explained by the fact that the energy density is dominated by the particles, with an almost negligible field energy density contribution of $\bm{E}^2/2 = E_L^2/2=E_\eta^2/(2\tau)$ which, at $\tau{\,=\,}\tau_0$, corresponds to 0.7\% of the particles' energy density $48\pi k T^4$. On the other hand, the pressure anisotropy $\pres_L/\pres_T$ shown in Fig.~\ref{F1}b exhibits a much stronger sensitivity to the interaction of the charged particles with the electric field. This interaction accelerates the isotropization of the pressure, most prominently for intermediate values of the specific shear viscosity $\bar\eta$.

Compared to the study performed in Ref.\ \cite{Ryblewski:2013eja} where a similar set of equations was investigated, the effect of the electric field on the pressure anisotropy is significantly reduced in our study. This is due in part to the smaller electric fields considered here, but the lack of a Schwinger term for the spontaneous decay of the fields in our treatment also contributes to this difference. As already stated we neglect the latter (and the emission of photons by the accelerated charges) since their rates are suppressed relative to strong-interaction collisions by the smallness of the electromagnetic coupling constant $\alpha=1/137$.

Proceeding to the electric current shown in Fig.~\ref{F1}c one observes at early times that the normalized diffusion current $J_L/T^3$ increases with increasing $\bar \eta$ (i.e. for larger collisional mean free path), but that for smaller $\bar\eta$ (i.e. larger microscopic collision rate) it persists longer. For $4\pi \bar \eta = 100$, i.e. close to the free streaming limit, the electric current is large enough to eventually flip the sign of the electric field, as seen in Fig.~\ref{F1}d. For the initial conditions chosen here this happens about 7\,fm/$c$ after starting the simulation. This reversal of the direction of the electric field repeats at larger times, i.e. over long time scales the electric field oscillates in time. This will be further explored in Fig.~\ref{F4} below.

%
\begin{figure*}[t]
\centering
\includegraphics[angle=0,width=0.75\linewidth]{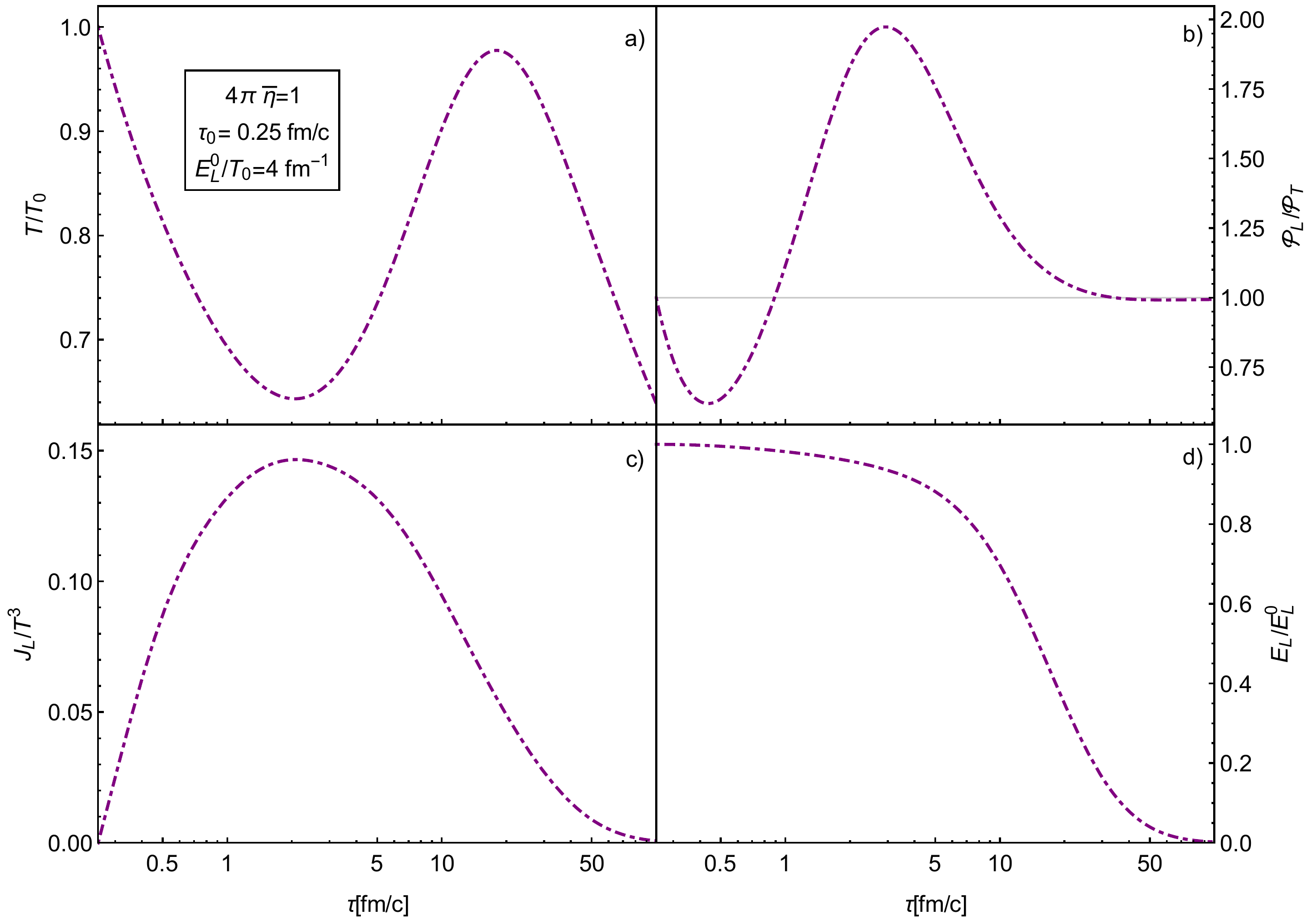}
\caption{ 
	(Color online) Long-time behavior of the evolution of the macroscopic quantities shown 
	in Fig.~\ref{F2}, for initial conditions $T_0=0.3$\,GeV and $E_L^0/T_0=4$\,fm$^{-1}$ at 
	$\tau_0=0.25$\,fm/$c$. Color coding and specific shear viscosity are the same as in Fig.~\ref{F2}.
	}
\label{F3}
\end{figure*}
%

For later comparison with the resummed moment expansion we also studied the exact solution for more extreme cases, corresponding to much larger initial electric fields ($\times 20$) and larger initial expansion rates ($\times 4$). They are shown in Fig.~\ref{F2}. To increase the initial expansion rate $\theta_0=1/\tau_0$ we simply decrease $\tau_0$ from 1 to 0.25\,fm/$c$. This entails a larger initial Knudsen number $\mathrm{Kn}^0\equiv\taueq^0 \, \theta_0=\taueq^0/\tau_0$ which increases from $\mathrm{Kn}^0 \lesssim 1/3$ to $\mathrm{Kn}^0 \gtrsim 1$ as we decrease $\tau_0$ from 1 to 0.25\,fm/$c$. Typically, non-hydrodynamic behavior is expected to occur when the Knudsen numbers are not sufficiently small. 

Figure~\ref{F2}a shows that changing the Knudsen number affects the temperature evolution only quantitatively (the system cools faster as the expansion rate and Knudsen number increase) but not qualitatively. When increasing the initial electric field by a factor 20, however, the temperature starts to evolve non-monotonically: the initial cooling stage is soon (after about 1.5-2\,fm/$c$) followed by an extended reheating period where Ohmic heating (i.e. the dissipation of the initially large electric field due to collisions among the charged particles) causes the temperature to increase again, in spite of the longitudinal expansion. 

The evolution of the pressure anisotropy shown in Fig.~\ref{F2}b exhibits even stronger sensitivity to the initial expansion rate and, in particular, the initial electric field. The reheating seen in Fig.~\ref{F2}a for the cases with large initial electric fields is accompanied (in fact, preceded) by a reversal of the pressure anisotropy from $\pres_L/\pres_T{\,<\,}1$ to $\pres_L/\pres_T{\,>\,}1$. Fig.~\ref{F2}c shows that the normalized electric current $J_L/T^3$ increases with both the initial electric field and initial expansion rate. Without the normalization by $T^3$, the non-monotonic behavior of $T(\tau)$ seen in panel (a) actually causes $J_L$ to oscillate, too, for large initial electric fields (not shown in Fig.~\ref{F2}). Only the electric field, shown in Fig.~\ref{F2}d, decreases continuously, without exhibiting any non-monotonic effects on the same time scales. Reheating, reversal of the pressure anisotropy and oscillations in the magnitude of the electric current do not occur for the smaller of the two initial values for the electric field studied here, irrespective of the initial expansion rate. 

A careful comparison between Figs.~\ref{F1}d and \ref{F2}d leads to an interesting observation: While increasing the specific shear viscosity $4\pi\bar\eta$ at fixed $\tau_0$ (i.e. fixed initial expansion rate) (Fig.~\ref{F1}) and increasing the initial expansion rate (by reducing $\tau_0$) at fixed $4\pi\bar\eta$ (Fig.~\ref{F2}) {\em both} increase the initial value of the Knudsen number $\mathrm{Kn}^0$, they have opposite effects on the initial rate of decrease of the electric field. Fig.~\ref{F1}d shows that $E_L$ initially decreases {\em more rapidly} when $4\pi\bar\eta$ is increased at fixed $\tau_0$ whereas we see in Fig.~\ref{F2}d that $E_L$ initially decreases {\em more slowly} when the initial expansion rate is increased at fixed $4\pi\bar\eta$, for both choices of the initial electric field value. Furthermore, we note from Fig.~\ref{F2}d that, at fixed initial expansion rate, the electric field initially decreases relatively more slowly, but at later times relatively more quickly as the initial value $E_L^0/T_0$ is increased. These observations reflect an interesting interplay between the two transport mechanisms related to the shear stress and charge diffusion currents that control the dissipative effects in this theory. 

%
\begin{figure*}[t]
 \centering
 \includegraphics[angle=0,width=0.8\textwidth]{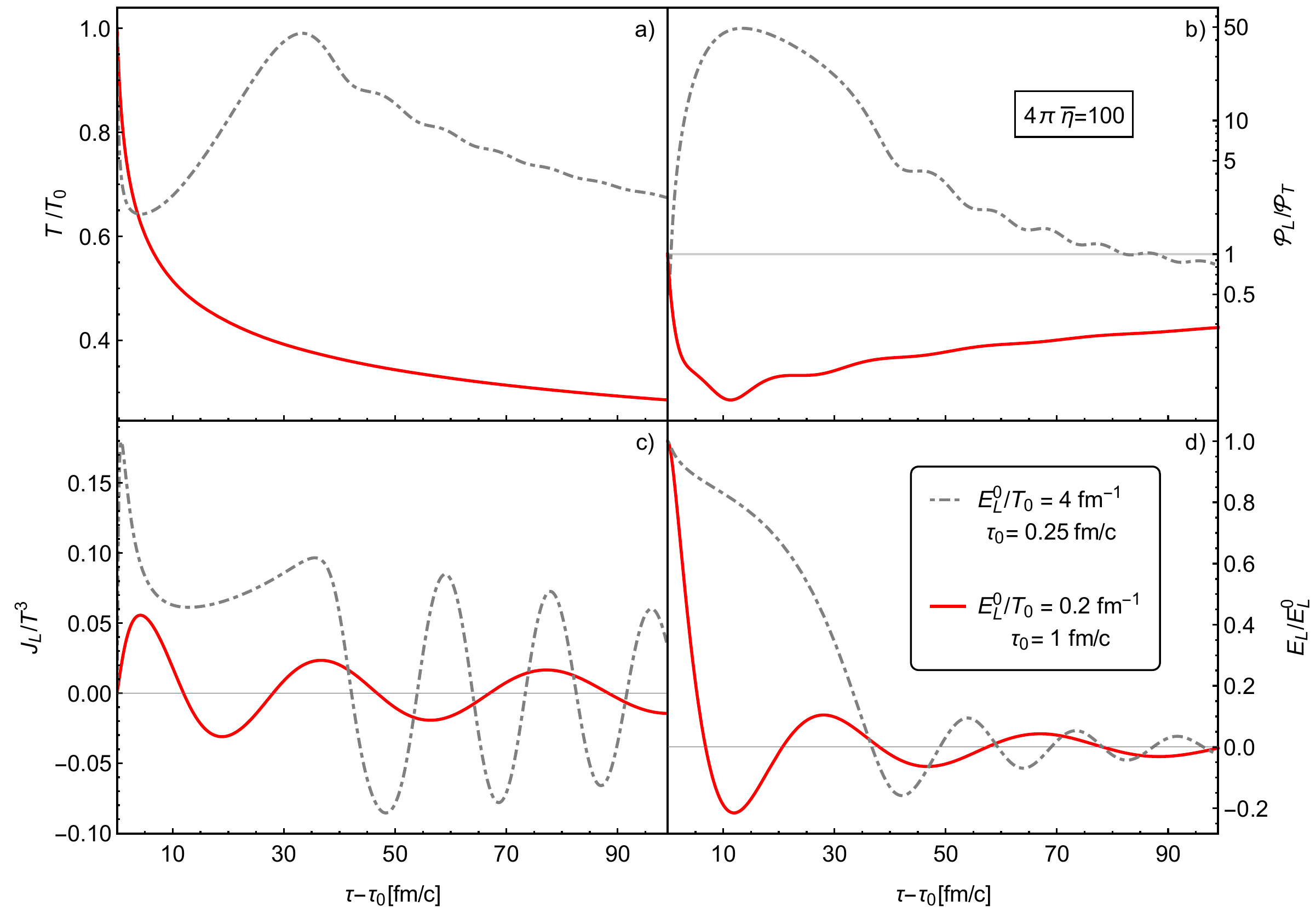}
\caption{
        Long-time evolution of (a) temperature $T/T_0$, (b) pressure anisotropy $\pres_L/\pres_T$,
        (c) electric current $J_L/T^3$, and (d) electric field $E_L/E_L^0$ for large shear viscosity 
        $4\pi\bar\eta=100$. 
	{\sl Red solid lines:} Same as the red solid lines in Fig.~\ref{F1}, but evolved over longer times. 
	{\sl Grey dash-dotted lines:} Similar, but for $20\times$ larger initial electric field and $4\times$ 
	larger initial expansion rate. 
	}
\label{F4}
\end{figure*}
%

We point out that for the larger of the two initial electric field values studied here the system is always far away from the Navier-Stokes limit. In order to see the ultimate approach to local thermal equilibrium we show in Fig.~\ref{F3} the long-time behavior of the macroscopic quantities studied in Figs.~\ref{F1} and \ref{F2}, for the most extreme case with both large initial electric field $E_L^0/T_0{\,=\,}4$\,fm$^{-1}$ and large initial expansion rate $\theta_0{\,=\,}4$\,fm$^{-1}c$. Clearly the system remains far from equilibrium up to 50\,fm/$c$ when the pressure ratio $\pres_L/\pres_T$ finally returns to values close to $1$. Beyond that time, the temperature, electric field, and electric current approach their asymptotic limits as power laws.\footnote{%
	We checked this by plotting Fig.~\ref{F3} double-logarithmically. We note that the current and 
	electric field approach their asymptotic power laws sooner than the temperature.
	}
In this regard, it would be interesting to study how the electric field affects the existence of hydrodynamic attractor solutions \cite{Heller:2015dha,Florkowski:2017olj,Strickland:2017kux,Denicol:2017lxn,Romatschke:2017vte} for the BVM equations. 
	
We already mentioned in the discussion of Fig.~\ref{F1} that, for very large mean free paths, $4\pi\bar\eta\to\infty$, corresponding to the collisionless Vlasov limit, the electric field flips sign at rather early times, hinting at long-time oscillations. This is further explored in Fig.~\ref{F4} where we consider $4\pi\bar\eta=100$ and show the behavior of the solid red lines in Fig.~\ref{F1} over a larger time interval $\tau-\tau_0=100$\,fm/$c$, together with a similar case with $20\times$ larger initial electric field and $4\times$ larger initial expansion rate. One sees that for large collisional relaxation times both the electric current and electric field (panels (c) and (d)) exhibit prominent long-time oscillations, even in the absence of a Schwinger term describing spontaneous electric field decay \cite{Ryblewski:2013eja}. For the moderate initial value of the electric field selected, $E_L^0/T_0=0.2$\,fm$^{-1}$ (see red solid lines), these field oscillations have only a minor effect on the evolution of the pressure anisotropy (shown in panel (b)) and leave no visible trace in the evolution of the temperature (panel (a)). This is, as before, easily explained by the small overall contribution of the electric field to the total energy density which is strongly dominated by the particle contribution. For much larger initial electric fields (grey dash-dotted lines), however, the oscillating energy content stored in the electric field and current is large enough to cause, through Ohmic heating, oscillations even in the temperature (a) and pressure ratio (b).

\subsection{Resummed moment equations for BVM with Bjorken expansion}
\label{sect:resummed}

Since our plasma contains both particles and antiparticles we must consider two generations of resummed moments,  $\gen^{\muif 1s}_r$ for the particles and $\bar \gen^{\muif 1s}_r$ for the antiparticles. The symmetry of the expansion reduces the number of independent degrees of freedom: Because of homogeneity in the transverse plane, moments with an odd number of $x$ or $y$ indices vanish - only pairs (or, more generally, even numbers) of $x$ and $y$ indices yield nonzero moments. In particular,
 \begin{equation}
 \begin{split}
  \gen_1^{xx{\muif 1s}} 
  & = \gen_1^{yy {\muif 1s}} = \int_p p_x^2 \cdots f = \frac{1}{2}\int_p p_T^2 \cdots f 
  \\
  & = \frac{1}{2}\int_p \,\Bigl[ (p\cdot u)^2 - \frac{p_\eta^2}{\tau^2} \Bigr] \cdots f 
  \\
  & = -\frac{1}{2}\Bigl( \partial_\xi^2 \gen^{\muif 1s}_1 + \frac{1}{\tau^2} \gen_1 ^{\eta\eta\muif 1s}\Bigr).
\end{split}
\end{equation}
With this result we can express all of the non-vanishing moments with $x$ and $y$ indices through moments of equal or lower rank with only $\eta$ indices. 

At this point it is convenient to use the longitudinal vector $z^\mu$ with Cartesian components $(\sinh\eta,0,0,\cosh\eta)$ or Milne components $(0,0,0,1/\tau)$ as it allows us to express the factor $p_\eta/\tau$ covariantly as $p_\eta/\tau = p\cdot z$. We use it to introduce the independent scalar moments
\begin{equation}\label{gen_pm}
 \begin{split}
  \gen_l^\pm 
  &\equiv \gen_1^{\muif 1l} z_{\mu_1}\cdots z_{\mu_l} 
        \pm \bar \gen_1^{\muif 1l} z_{\mu_1}\cdots z_{\mu_l}
  \\
  &= \int \frac{d^3 \tilde p}{\tau} \, \left(\frac{p_\eta}{\tau}\right)^l e^{-\xi^2\left( p_T^2 + p_\eta^2/\tau^2 \right)} 
        \Bigl(f\pm \bar f \Bigr).
 \end{split}
\end{equation}
For the Bjorken expanding case, the scalars $\{\gen^\pm_0,\cdots,\gen^\pm_l\}$ contain all the information about the resummed moments up to tensor rank $l$ of the system. Their exact evolution can be obtained directly from Eq.~(\ref{gen_ev_BV}), taking $l$ projections along $z$ and summing (subtracting) the particle and anti-particle equations. An equivalent method consists in taking directly the $\tau$ derivative of the right-hand side of Eq.~(\ref{gen_pm}). Either way one finds
\begin{equation}\label{0+1_moments_evolution}
 \begin{split}
  & \partial_\tau \gen^\pm_l +\frac{1}{\taueq}\left( \gen^\pm_l -\gen^\pm_{l,\mathrm{eq}} \right) = \\
  & -\frac{l+1}{\tau} \gen^\pm_l +\frac{2\xi^2}{\tau}\gen^\pm_{l+2} + \frac{q E_\eta}{\tau}\left[\vphantom{\frac{}{}} l\, \gen^\mp_{l-1} - 2\xi^2 \gen^\mp_{l+1} \right]. 
 \end{split}
\end{equation}
In the last term of Eq.~(\ref{0+1_moments_evolution}) it is understood that for $l=0$ the first term in the square brackets is zero.

The equilibrium moments $\gen^{-}_{l,\mathrm{eq}}$ vanish for all $l$ while the equilibrium $+$-moments $\gen^{+}_{l,\mathrm{eq}}$ vanish for odd $l=2n{+}1$. For even $l=2n$ one has
\begin{equation}
\label{gen_eq}
  \gen^+_{2n,\mathrm{eq}}  =  \frac{8\pi k}{T}\, \frac{(2n{+}2) (2n)!}{(2\xi )^{2n+4}} \
  U\!\left(2{+}n,\frac{3}{2},\frac{1}{(2\xi T)^2}\right)
\end{equation}
where $U(a,b,z)$ is the Tricomi confluent hypergeometric function \cite{NIST:DLMF}.

Full knowledge of the hydrodynamic moments (i.e. the stress-energy tensor and the electric current) is encoded in the three scalar moments $\gen^+_0$, $\gen^-_1$, and $\gen^+_2$. Therefore, at leading order in the resummed moment expansion only the $\gen^\pm_l$ up to $l=2$ are considered as dynamical variables, i.e. we truncate the hierarchy of moment equations (\ref{0+1_moments_evolution}) at $l=2$. As higher order corrections we will progressively consider the higher order moments as additional dynamical variables.\footnote{%
		Because of symmetry, the odd $\gen^+_{2n+1}$ and the even $\gen^-_{2n}$ never couple directly 
		to any of the other moments. Their evolution couples with the hydrodynamic moments only indirectly
		through the electric field, which is itself coupled to the electric current. In particular, if the initial 
		values of all these moments are zero (for instance for local equilibrium initial conditions) their 
		evolution is trivial. 
		} 
%

%
\begin{figure*}[t!]
\centering
\includegraphics[angle=0,width=0.79\linewidth]{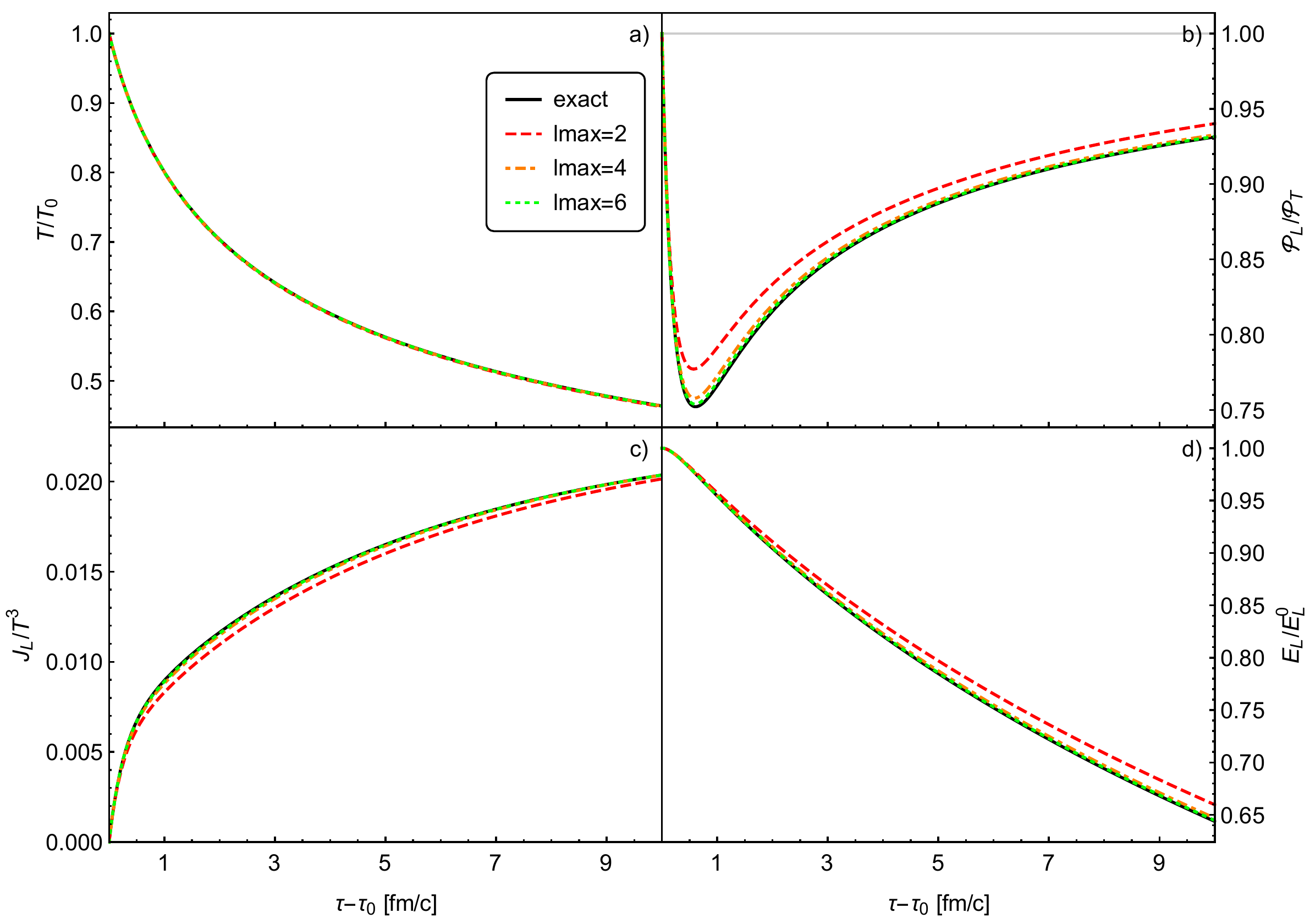}
\caption{
	(Color online) Comparison of the exact time evolution (black solid lines) with that from the resummed 
	moment expansion truncated at different orders $l_\mathrm{max}$ as specified in the legend (colored 
	lines), for the temperature $T/T_0$ (a), pressure ratio $\pres_L/\pres_T$ (b), electric current $J_L/T^3$
	(c), and electric field $E_l/E_L^0$ (d), evolved with minimal specific shear viscosity $4\pi\bar\eta=1$. 
	Initial conditions are $T_0=0.3$\,GeV and $E_L^0=0.2$\,fm$^{-1}$ at $\tau_0=1$\,fm/$c$.  
	}
\label{F5}
\end{figure*}
%

When truncating the moment hierarchy at some maximum $l$ value $l_\mathrm{max}$, there is no unique prescription to approximate the moments of orders $l_\mathrm{max}{+}1$ and $l_\mathrm{max}{+}2$ appearing on the r.h.s. of Eq.~(\ref{0+1_moments_evolution}). The simplest one is to assume $\gen^\pm_l \approx \gen^\pm_{n,\mathrm{eq}}$ for $n{\,>\,}l_\mathrm{max}$. For numerical purposes this approximation is very helpful because we can use the simple formula (\ref{gen_eq}) to evaluate the non-dynamical higher-order moments. It must be noted, though, that this approximation is not the most accurate possible. Recall that the tensor moments $\gen^{\muif 1s}_r$ are defined on a non-orthonormal polynomial basis of tensors constructed from the locally spatial momentum vectors $p^{\langle\mu\rangle}$. Due to the lack of orthogonality, some information on the higher-order moments is already contained in the lower order ones and this could be exploited.\footnote{%
		The situation is analogous the one presented in~\cite{Denicol:2016bjh} for the (non-resummed) reducible
		moments.
		} 
We will here ignore this extra information, leaving the most effective approximation scheme for the non-dynamical higher-order moments for future research.

It is numerically convenient to normalize the scalar projections $\gen^\pm_l$ in order to confront pure numbers of similar order of magnitude in the evolution equations. For even $l=2n$ we do so by dividing them by the $\xi\to0$ limit of their equilibrium values: 
\begin{equation}
   \lim_{\xi\to 0}\gen^+_{2n} = 8\pi k (2n{+}2)(2n)! \,T_0^{2n +3}.
\end{equation}
The right hand side can be extended to the case of odd $l$, and we can use the same normalizing factors for both plus and minus moments:
\begin{equation}\label{norm_mom}
  M^\pm_{l} \equiv \frac{\gen^\pm_l}{8\pi k(l{+}2)\, l! \,  T_0^{l+3}}.
\end{equation}
Obtaining from Eq.~(\ref{0+1_moments_evolution}) the evolution of the normalized moments (\ref{norm_mom}) is straightforward:
\begin{eqnarray}
\label{norm_mom_ev}
   &&\partial_\tau M^\pm_l +\frac{1}{\taueq}\left( M^\pm_l -M^\pm_{l,\mathrm{eq}} \right) 
   \nonumber\\
   &&= -\frac{l{+}1}{\tau} M^\pm_l +\frac{2(\xi T_0)^2}{\tau}(l{+}4)(l{+}1)M^\pm_{l+2} 
   \\\nonumber
   &&\quad \ + \frac{q E_\eta}{\tau \, T_0}
         \left[  \frac{l{+}1}{l{+}2} M^\mp_{l-1} - 2(\xi T_0)^2 \frac{(l{+}3)(l{+}1)}{l{+}2}M^\mp_{l+1} \right].
\end{eqnarray}
It must be noted that for $l =0$ the (otherwise undefined) first term $\sim M^\mp_{-1}$ inside the square brackets is absent.

%
\begin{figure*}[t]
\centering
\includegraphics[angle=0,width=0.8\linewidth]{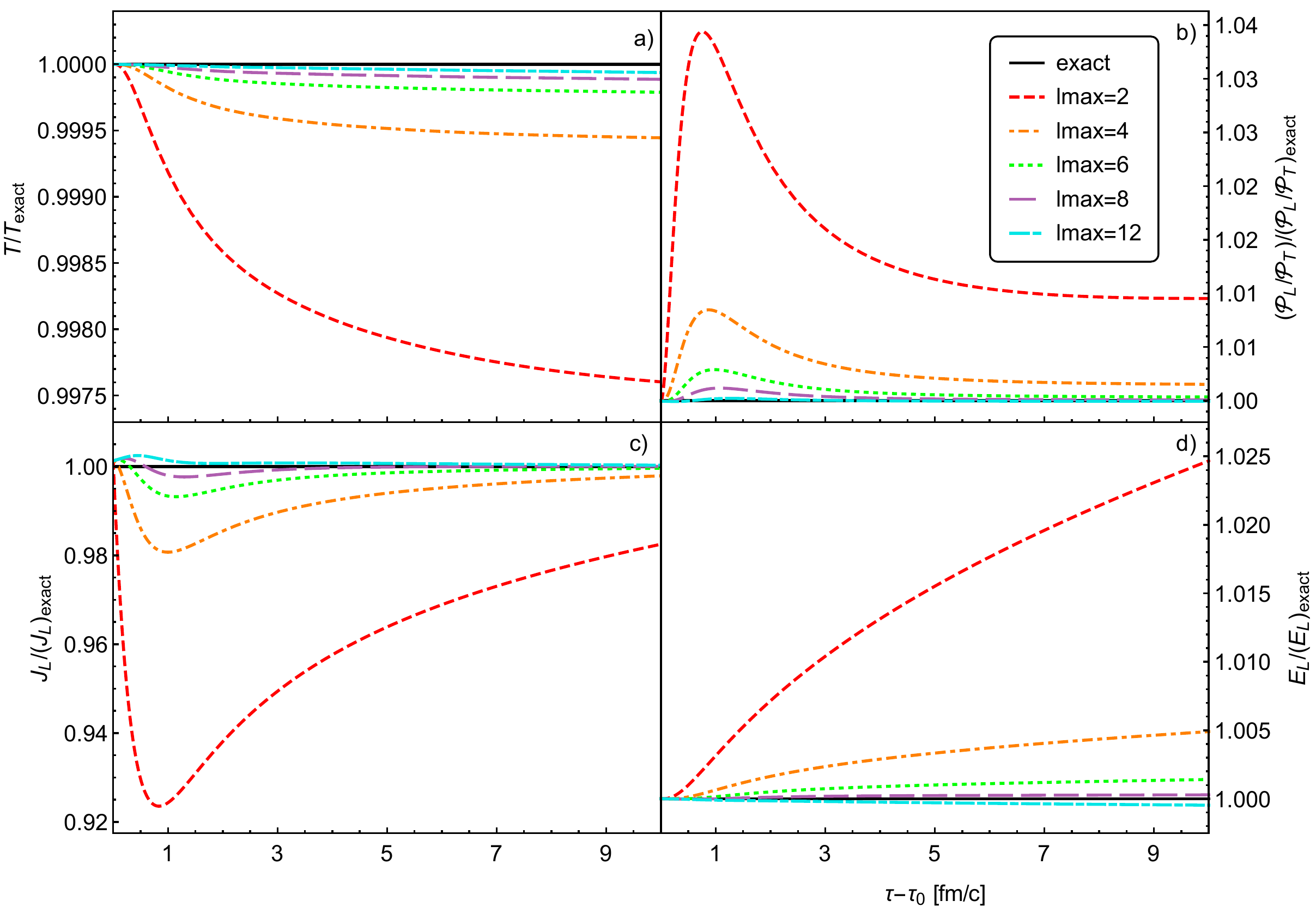}
\caption{
	(Color online) Similar to Fig.~\ref{F5}, but here presented as relative deviations from the exact results 
	of the results from the resummed moment expansion for different truncation orders up to 
	$l_\mathrm{max}=12$. 
	}
	\label{F6}
\end{figure*}
%

The effective temperature is defined through the Landau matching procedure (\ref{exact_temp}). Making use of the normalized moments, the effective temperature can be written as 
\begin{eqnarray}
\label{eq55}
  48 \pi k T^4 &=& \int_p (p\cdot u)^2 \,\bigl( f{+}\bar f \bigr)
  = \frac{2}{\sqrt{\pi}} \int_0^\infty d\xi \bigl(-\partial_{\xi^2}\gen_0^+ \bigr)\nonumber\\
 & =& 32\sqrt{\pi} k T_0^3  \int_0^\infty d\xi \bigl(-\partial_{\xi^2}M^+_0 \bigr)
\end{eqnarray}
or, equivalently,
\begin{equation}
\label{eq56}
  T = \Bigl[ \frac{2}{3\sqrt{\pi}}T_0^3 \int_0^\infty d\xi \bigl(-\partial_{\xi^2} M^+_0 \bigr) \Bigr]^{\frac{1}{4}}.
\end{equation}
Implementing Landau matching is numerically expensive since the temperature must be matched to the energy density moment by evaluating the integral on the r.h.s. of (\ref{eq56}) at each time step. It turns out to be more convenient to instead treat $T$ as an additional dynamical variable. Its exact evolution can be obtained by differentiating the r.h.s. of Eq.~(\ref{eq56}) with respect to time and inserting Eq.~(\ref{norm_mom_ev}) for $l=0$. 
One finds
\begin{eqnarray}
\label{eq57}
    &&\partial_\tau T =  -\frac{1}{4\tau}\bigg[T + \frac{1}{24(\pi)^{\frac{3}{2}} k T^3}\int_0^\infty \!\! d\xi \, \phi^-_2 
\nonumber\\    
    && \qquad\qquad\qquad\  -\, \frac{q E_\eta}{24(\pi)^{\frac{3}{2}} k T^3}\int_0^\infty\!\! d\xi \, \phi^-_1 \bigg] 
\\\nonumber
    && =  -\frac{1}{4\tau}\left[ T + \frac{8 T_0^5}{3\sqrt{\pi} T^3}\int_0^\infty \!\! d\xi \, M^-_2 
    - \frac{q E_\eta}{\sqrt{\pi} T^3}\int_0^\infty\!\! d\xi \, M^-_1 \right].
\end{eqnarray}

The last equation to solve numerically is the evolution of the electric field (\ref{exact_el}). Written in terms of the normalized moments it becomes
\begin{equation}
 \begin{split}
  \partial_\tau E_\eta 
  &= \frac{1}{\tau} E_\eta - \frac{2 q \tau}{\sqrt{\pi}}\int_0^\infty\!\! d\xi\, \gen_1^- 
  \\
  &=\frac{1}{\tau} E_\eta - 48 k  q  \tau T_0^4 \sqrt{\pi} \int_0^\infty \!\! d\xi\, M_1^-.
 \end{split}
\end{equation}
These last two equations, coupled to Eqs.~(\ref{norm_mom_ev}) for those moments that we treat dynamically, can be solved with an explicit Runge-Kutta method. In the following subsection we compare the evolution of the macroscopic quantities (i.e. the temperature, pressure anisotropy, electric current, and electric field) between the solution of the resummed moment equations and the exact solution of the Boltzmann-Vlasov-Maxwell equations discussed in the previous subsection. 

\subsection{Comparison of the resummed moment expansion with the exact solution of the BVM equations}
\label{sect:comp}

%
\begin{figure*}[t]
\centering
\includegraphics[angle=0,width=0.48\linewidth]{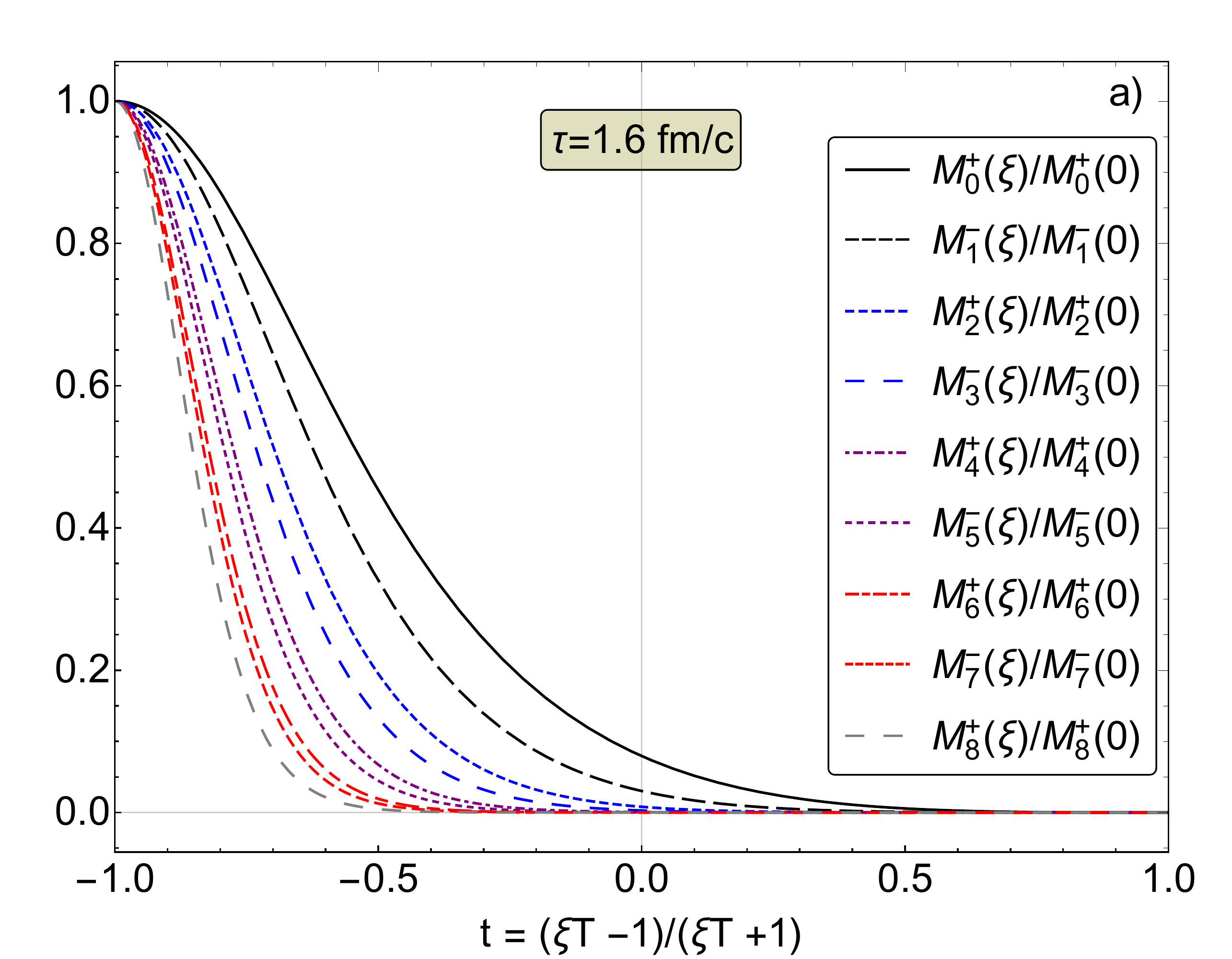}
\includegraphics[angle=0,width=0.48\linewidth]{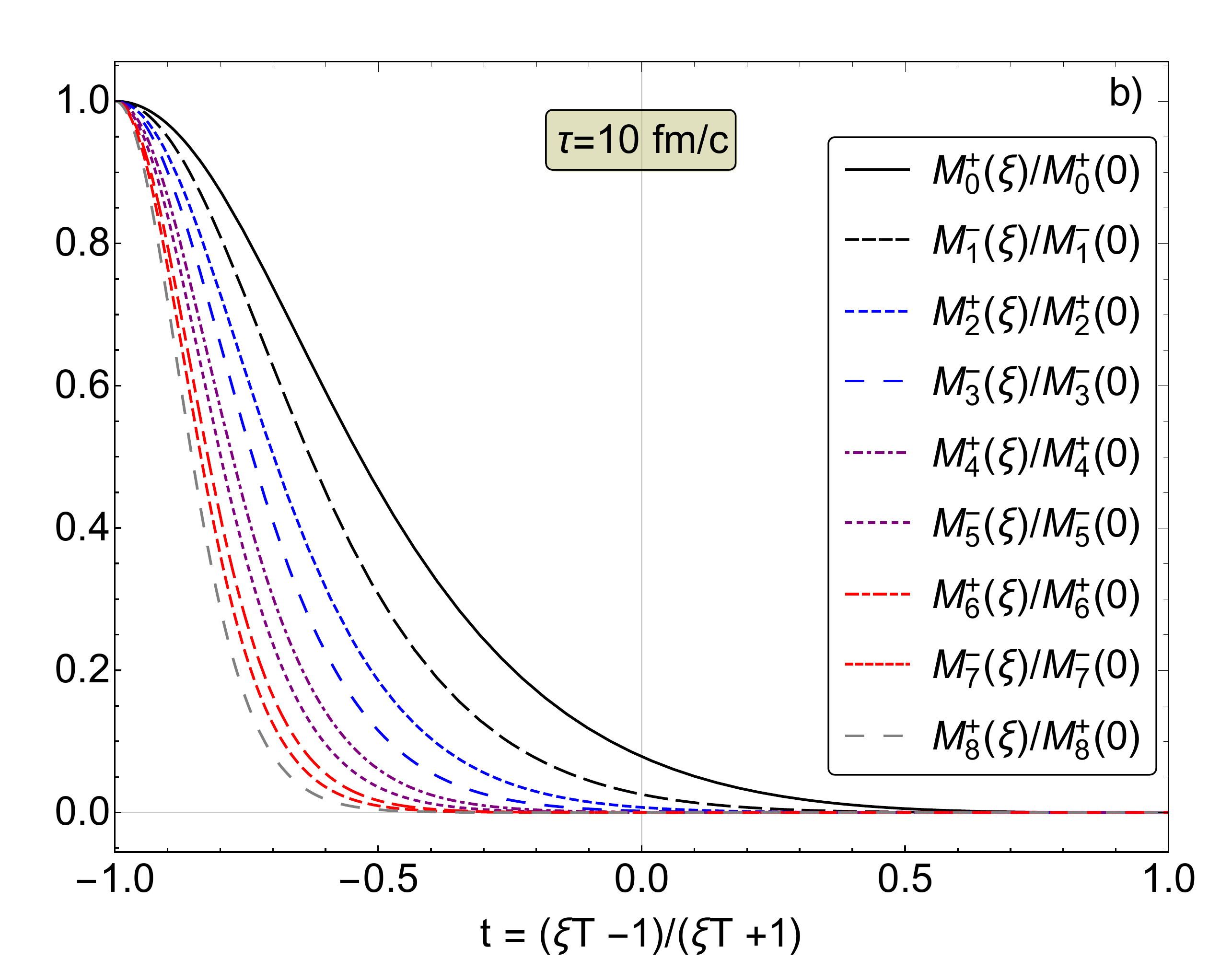}
\caption{
	(Color online) The resummed moments $M^\pm_l$ for $l=1,\dots,8$, normalized by their values 
	at $\xi{\,=\,}0$, as functions of $t\equiv(\xi T{-}1)/(\xi T{+}1)$, at two proper time values $\tau=1.6$ 
	(left panel) and 10\,fm/$c$ (right panel). Evolution parameters are the same as in Fig.~\ref{F5}. The 
	moments $M^\pm_l$ are seen to decay to zero over the natural length scale $\xi\sim 1/T$. (Note that 
	the temperature $T$ 	drops by more than 1/3 between $\tau=1.6$ and 10\,fm/$c$, see Fig.~\ref{F5}a.)
	}
\label{F7}
\end{figure*}
%

As an example for studying the convergence of the resummed hydrodynamic moment expansion, we consider the default initial conditions used for the exact solutions shown in Fig.~\ref{F1} ($T_0{\,=\,}0.3$\,GeV and $E_L^0/T_0{\,=\,}0$ or 0.2\,fm$^{-1}$ at $\tau_0{\,=\,}1$\,fm/$c$), with minimal shear viscosity $4\pi\bar\eta{\,=\,}1$ (black lines). In Fig.~\ref{F5} we compare these exact solutions with three iterations of the resummed moment expansion, corresponding to truncations at different $l_\mathrm{max}$ values as described in the preceding section, i.e. with thermal equilibrium values $M^\pm_{n,\mathrm{eq}}$ for all moments of order $n>l_\mathrm{max}$. In Fig.~\ref{F5} the results from the resummed moment expansion are nearly indistinguishable from the exact results already for $l_\mathrm{max}{\,=\,}6$. Fig.~\ref{F6} shows that a relative precision better than $1/1000$ is achieved with $l_\mathrm{max}{\,\geq\,}8$. We verified the continued convergence of the series up to $l_{\rm max}=34$ where we stopped the calculation.

Figure~\ref{F6} shows that for $\tau{-}\tau_0{\,<\,}10$\,fm/$c$ the resummed moment expansion agrees with the exact solution to better than 10\% already at leading order $l_\mathrm{max}{\,=\,}2$, i.e. in the hydrodynamic limit. The largest deviation is seen for the current $J_L/T^3$ where it reaches about 8\%, followed by 3.5\% for the pressure ratio and 2.5\% for the electric field. We found similar levels of (im)precision for all other parameter sets studied in this work. Larger deviations from the exact result at leading order are accompanied by slower convergence towards the exact result as the order of the approximation is increased. But in most cases $l_\mathrm{max}{\,=\,}12$ resulted in relative deviations from the exact result of less than 1\%. Large Knudsen numbers (large expansion rates) and large initial pressure anisotropies have surprisingly weak effects on the speed of convergence. 

A significant slowing of the rate of convergence was noted only for very large initial electric fields and/or large relaxation times (i.e. large specific shear viscosities, close to the free-streaming limit). In both cases the reason for slower convergence is easily understood by inspecting Eq.~(\ref{norm_mom_ev}). There are only two couplings to moments of higher orders, one of which is multiplied by the electric field. No matter the order of the approximation, in the equation for $l{\,=\,}l_\mathrm{max}$ the moments $M^\mp_{l_\mathrm{max}+1}$ and $M^\pm_{l_\mathrm{max}+2}$ must be approximated. A large electric field, or better yet, a large (dimensionless) ratio $E_\eta(\tau)/T_0$, magnifies the coupling to one of the approximated moments. The approximation error then affects the evolution of all the lower moments since they all couple to the higher ones via the electric field. For large relaxation times one must recall that the temperature decreases more slowly than the expansion rate, due to Ohmic heating. Normally the second term on the left hand side of Eq.~(\ref{norm_mom_ev}), which is proportional to $1/\taueq{\,=\,}T(\tau)/(5\bar\eta)$, soon dominates the evolution since the scalar expansion rate $\theta =1/\tau$ suppresses all terms on the right hand side. This stage is, however, postponed as $\bar\eta\to\infty$. In the free-streaming limit the right hand side, including the terms coupling to the higher order moments, becomes dominant, and thus the evolution becomes more sensitive to  approximations made for the moments of order $n>l_\mathrm{max}$.

While illuminating, these considerations are not enough to understand the reasons for the observed generically rapid convergence of the resummed moment expansion, nor do they throw light on possible situations that might spoil this convergence when going from the simple case of Bjorken flow to a more realistic 3-dimensional expansion. It is worth noting that nowhere in the discussion of the convergence of the resummed moment expansion is the influence of large  gradients or large pressure corrections mentioned. The present formalism thus appears unrelated to the traditional expansion in powers of Knudsen and inverse Reynolds numbers \cite{Denicol:2012cn} which have been used to define the effective range of validity of relativistic hydrodynamics. Within the resummed moment expansion, the only approximation lies in the coupling to moments of order $n>l_\mathrm{max}$, and only these (non-hydrodynamic) moments are approximated (in our case with the assumption of small deviations from equilibrium\footnote{%
	Perhaps better approximations for the part of the $n>l_\mathrm{max}$ 
	moments that is orthogonal to the lower moments can be found.}%
).

%
\begin{figure*}[t]
\centering
\includegraphics[angle=0,width=0.8\linewidth]{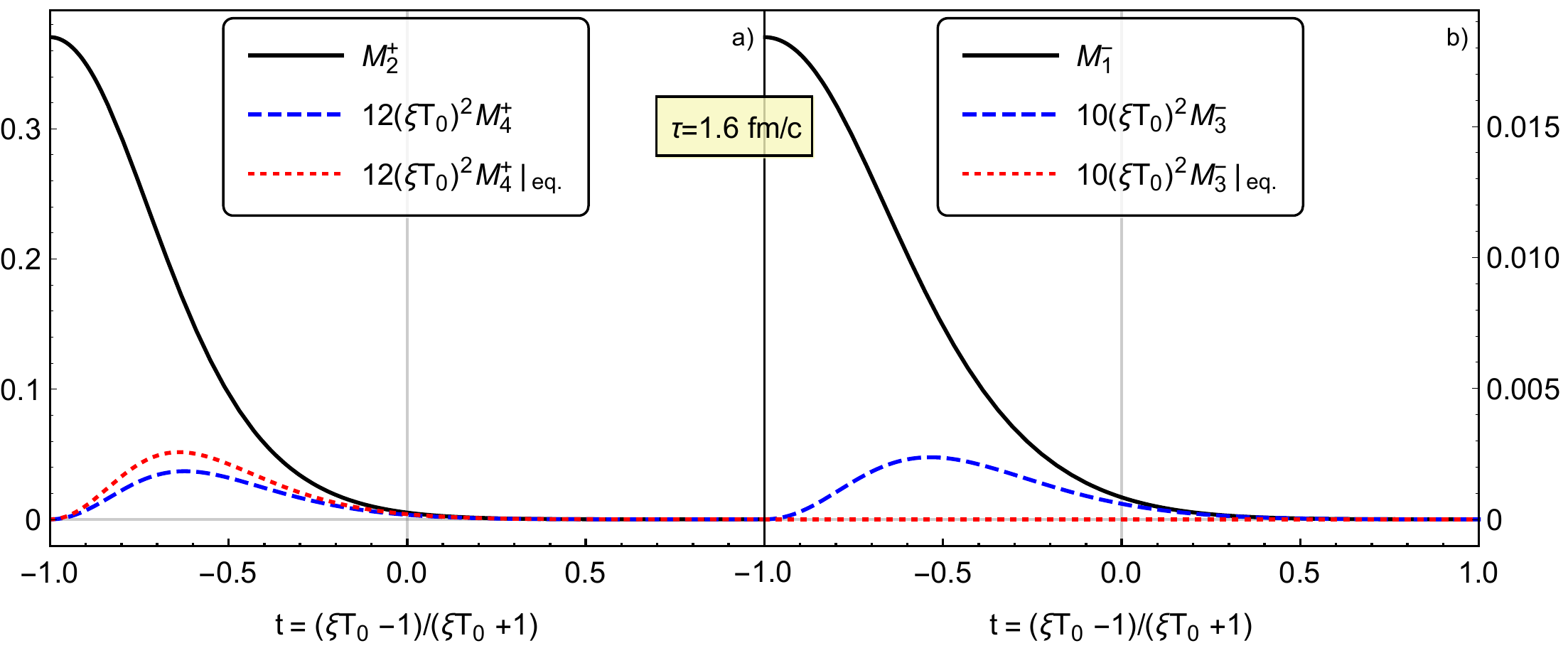}
\caption{
	(Color online) The left and right terms in the first (a) and second lines (b), respectively,
	of the r.h.s. of Eq.~(\ref{norm_mom_ev}) for $l{\,=\,}2$, plotted at $\tau{\,=\,}1.6$\,fm/$c$ as
	functions of $t\equiv(\xi T_0{-}1)/(\xi T_0{+}1)$. 	
	}
\label{F8}
\end{figure*}
%

So how does approximating these higher-order, non-hydrodynamic moments affect the dynamical evolution of the hydrodynamic moments, i.e. of the charge current and energy-momentum tensor? The following observations may bring us closer to an understanding of the good convergence of the moment expansion even in situations where traditional arguments based on the size of hydrodynamic gradients and/or dissipative flows (i.e. on Knudsen and/or inverse Reynolds numbers) suggest the breakdown of hydrodynamics. Looking at Eq.~(\ref{norm_mom_ev}) one sees that the higher order moments always appear multiplied by a factor $(\xi T_0)^2$. A crucial observation is that, in the situations studied in this paper, the resummed moments are all strongly peaked around $\xi\simeq 0$, especially the higher order ones (see Fig.~\ref{F7}). This is because the resummed moments in Eq.~(\ref{resummed_Gen}), and hence their (normalized) $(0+1)$-dimensional versions in Eqs.~(\ref{gen_pm}) and (\ref{norm_mom}), contain a Gaussian factor $e^{-\xi^2(p\cdot u)^2}$ that effectively suppresses all integrals at large values of $\xi$. It is especially true for the higher-rank moments which, due to extra momentum factors in the integrand, probe the distribution function at higher momenta where the suppression by the Gaussian becomes effective already at lower values of $\xi$.

On the r.h.s. of Eq.~(\ref{norm_mom_ev}) we already grouped the terms (both those with and without an electric field factor) accordingly. In both lines the term with the lower-order dynamically evolved moment will dominate over the one involving the higher-order moment because, in the $\xi$ region where the moments are substantial, the latter is suppressed by a factor $\xi^2$. For $l=l_\mathrm{max}$, in particular, this means that the coupling to the moments of order $l_\mathrm{max}{+}1$ and $l_\mathrm{max}{+}2$ that are not evolved dynamically, but instead approximated by their equilibrium values, are suppressed. This suppression of the influence from higher-order moments repeats at each step of increasing the order $l$, which largely explains the observed rapid convergence of the resummed moment method.   

Fig.~\ref{F8} illustrates this mechanism for the $+$ moment of order $l{\,=\,}2$, $M^+_2$, which is responsible for the pressure anisotropy. The black solid lines show the $\xi$-dependence of the leading moments $M^+_2$ and $M^-_1$ in the upper and lower lines of the r.h.s. of Eq.~(\ref{norm_mom_ev}), plotted as functions of the transformed variable $t\equiv(\xi T_0{-}1)/(\xi T_0{+}1)$ whose origin lies at $\xi=1/T_0$. The blue-dashed lines show the subleading second terms in these two lines (the ones that describe the coupling to moments of higher order) as obtained from a calculation where all moments up to order $l_\mathrm{max}=24$ were evolved dynamically. The red-dotted lines show the effect of truncating the resummed moment expansion at leading order (i.e. at $l_\mathrm{max}{\,=\,}2$) and approximating the moments $M^+_4$ and $M^-_3$ by their thermal equilibrium values. In both panels the coupling terms to the higher-order moments are seen to be dwarfed by the leading terms over the entire $\xi$ range, but especially at $\xi{\,=\,}0$ ($t{\,=\,}{-}1$) where the resummed moments peak.

The suppression of coupling terms to higher order resummed moments is expected to partially persists in $3{+}1$ dimensions. It is difficult to think of a realistic phase-space distribution that will not produce resummed moments that peak near $\xi=0$ and decay for $\xi$ larger than the inverse of the typical energy scale of the system (which is normally larger than the temperature to which the system settles after thermalization). For massive particles a momentum-independent suppression factor $e^{-\xi^2 m^2}$ can be extracted from the defining integral~(\ref{resummed_Gen}). It is worth noting that the Gaussian suppression factors arise only in the resummed expansion; the ordinary expansion equations for the hydrodynamic moments are recovered from the resummed moments by performing a $\xi$ integral for the scalar, vector and rank two tensor equations.

Still, looking at the general equation (\ref{gen_ev_BV_r=1}) one notices two terms in the second integral on the r.h.s. that couple to moments of higher tensor rank without being suppressed by factors of $\xi^2$: the first one involving the acceleration and the last one which is a space-divergence. In the general case these terms have the potential of spoiling the rapid convergence of the resummed moment expansion and thus the good agreement of the leading-order hydrodynamic approximation with the exact solution of the Boltzmann-Vlasov-Maxwell equations. Future research will settle this issue. There is, however, a somewhat handwaving argument suggesting that the resummed moment expansion will indeed continue to converge in the general $(3{+}1)$-dimensional case. Looking again at Eq.~(\ref{gen_ev_BV_r=1}) one sees that \textit{at any order} only two higher order moments couple without $\xi^2$ suppression to the evolution of the lower ones (five if one includes the $\xi^2$-suppressed couplings). On the other hand, the number of moments of the same or lower rank increases rapidly with the tensor rank $s$. Arguing that all the moments of a given tensor rank and order have the same physical dimensions and are therefore expected to be of similar order of magnitude, exact cancellations between many of them would be needed for their contribution to the evolution equations to be overwhelmed by the two terms coupling to higher ranks or orders. This argument works both with and without coupling to the electromagnetic fields.

We note that a similar argument cannot be made for the traditional moment expansion, even in the absence of electromagnetic fields where the coupled equations couple only to moments that remain well defined and well behaved in the limit $m/T\ll1$. Equation~(\ref{red_ev}) for the evolution of the standard $\reducible$-moments contains $s{+}3$ terms coupling to moments of different order and only $s{+}1$ moments of the same tensor rank $s$ and energy index $r$. Moreover, two of the couplings to different moments are multiplied by $r$ and $r{-}1$. Recalling that all members of the tower of moments coupled to the hydrodynamic moments feature the same sum $r{+}s$, one must therefore expect that at the truncation step the evolution of the last dynamically evolved moments is dominated by the $\|r\| + \|r-1\| +1 +s$ approximated moments, unless the approximation is fine-tuned to render their combined contribution subleading relative to the $s{+}1$ dynamically evolved moments. Such a cancellation does not even happen automatically in the highly symmetric Bjorken case; therefore, even there one must put some effort into designing a good enough approximation scheme for the non-dynamical moments that does not ruin the convergence properties of the expansion.

%
\begin{figure}[t]
\centering
\includegraphics[angle=0,width= \columnwidth]{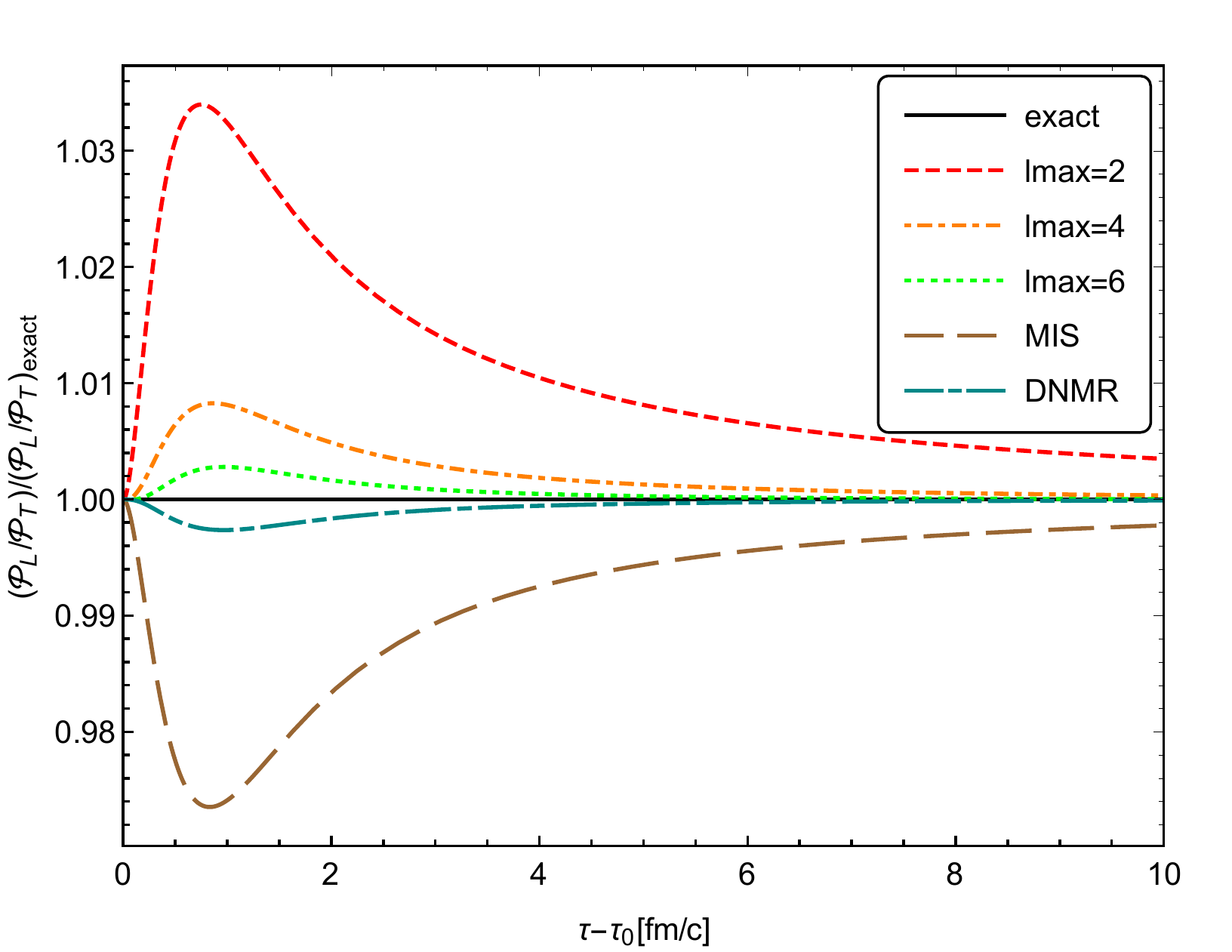}
\caption{
	(Color online ) Comparison between the resummed moment expansion at different truncation orders 
	(see legend) and two second-order viscous hydrodynamic approximations (MIS and DNMR), for the 
	pressure anisotropy in the absence of an electric field.
         }
\label{F9}
\end{figure}
%

We close this section with the following remark. In this work we used the approximation $M^\pm_n = M^\pm_{n,\mathrm{eq}}$ for the normalized moments of order $n>l_\mathrm{max}$ where $l_\mathrm{max}$ is the order of the last dynamically evolved moment. This corresponds in the general (3+1)-dimensional case to the approximation $\gen^{\muif 1 s}_r=\gen^{\muif 1 s}_r|_{\rm eq.}$. Due to its simplicity, this is a particularly convenient approximation if one wants to check the convergence properties up to very high orders. However, it does not make full use of all the information from the lower-order moments. This is illustrated in Fig.~\ref{F9} where we compare the precision of the resummed moment method in reproducing the exact pressure anisotropy obtained from the Boltzmann equation (without electric fields) at different orders of truncation with that of two well-established second-order dissipative hydrodynamic approximations for uncharged fluids that were derived using different methods, namely the M\"uller-Israel-Stewart (MIS) \cite{MIS-1,MIS-6} and Denicol-Niemi-Moln\'ar-Rischke (DNMR) theories \cite{Denicol:2012cn}. Fig.~\ref{F9} shows that at leading order the resummed moment approximation reaches almost the same level of precision as the MIS theory, albeit with deviations from the exact result that have opposite signs. The DNMR approximation, on the other hand, is significantly closer to the exact solution; with the resummed moment expansion, using our approximation scheme, one must go to $l_\mathrm{max}{\,=\,}6$ (i.e. NNLO) to reach comparable precision. 

At leading order without electric fields, the moment evolution equations for the resummed moment expansion and the traditional method of moments (DNMR) have the same form; they differ only in the approximation for the higher order moments. In particular, the DNMR approach does not assume thermal equilibrium values for the higher-order moments, but instead relates them to the dynamically evolved lower-order moments. One can imagine doing something similar in the resummed moment expansion. Other approximations can also be considered, such as anisotropic hydrodynamic \cite{Alqahtani:2017mhy} expectation values. We leave the problem of optimizing the approximation for the non-dynamical higher-order moments to future research.

\section{Conclusions}
\label{sect:conclusions}

In this paper we studied the emergence of hydrodynamics from kinetic theory for an overall electrically neutral system of charged particles described by the Boltzmann-Vlasov-Maxwell equations. We found that for this problem a straightforward generalization of the standard method of moments becomes considerably more intricate, especially in the massless limit. We pursued a different approach which consists of introducing an expansion in terms of a set of resummed moments that converges rapidly for systems both with and without long-range mean fields. As was the case for the standard moments, the components of the particle (charge) current and energy-momentum tensor (i.e. the hydrodynamic degrees of freedom of the system) can be expressed through certain low-order resummed moments. Truncating the hierarchy of coupled evolution equations for the resummed moments at lowest order yields a set of hydrodynamic equations. The precision of the description of the microscopic dynamics via moment equations can be systematically improved by truncating the hierarchy at higher order, thereby including the effects of non-hydrodynamic moments of the distribution function into the evolution of the hydrodynamic degrees of freedom. Therefore, the resummed moments method provides a convenient and reliable basis for the hydrodynamic expansion of a plasma interacting with long-range mean fields.

We studied the convergence of this expansion, and the precision of the resulting macroscopic hydrodynamic description as a function of the truncation order, for a highly symmetric expansion scenario (Bjorken flow), where symmetries (longitudinal boost invariance and transverse homogeneity) simplify the system of equations to the extent that the microscopic kinetic evolution defined by the BVM equations can be solved exactly. Exploring a range of model parameters such as the initial expansion rate (controlled by the starting time $\tau_0$), the initial value of the electric field, and the specific shear viscosity of the fluid (controlled by the relaxation time encoded in the RTA Boltzmann collision term), we found uniformly rapid convergence of the moment expansion to the exact solution. 

As a figure of merit one should remember that a precision of better than 1/1000 for the hydrodynamic variables is typically reached by truncating the resummed moment expansion at order $l_\mathrm{max}=8$ or higher. When performing the truncation, we here approximated the higher-order moments to which the last dynamically evolved moment couples by their thermal equilibrium values; though other approximation schemes are possible they were not investigated in this paper. The most efficient way to slow down the convergence of the expansion is to increase the viscosity to large values, i.e. to push the theory towards the free-streaming limit. Slower than typical convergence was also found for very large values of the initial electric field.\footnote{%
	For the extreme case shown as grey lines in Fig.~\ref{F4}, we were unable to achieve relative precision 
	of $10^{-4}$ even for $l_\mathrm{max}=34$, but we are not certain whether this indicates a limitation of 
	the precision of the numerical method used, or slow convergence of the resummed moment expansion.
	}  
The first of these two observations can be explained by observing that in Eq.~(\ref{norm_mom_ev}) the coupling of the moment $M^\pm_l$ to itself dominates over the coupling to the (approximated) higher moments as long as the inverse Knudsen number $\tau/\taueq$ is small, but that the evolution becomes very sensitive to the approximation of the higher moments in the opposite limit, i.e. for large relaxation times $\taueq$. The second observation is explained by the coupling of $M^\pm_l$ to $M^\mp_{l{+}1}$ (which at truncation level is approximated non-dynamically) and the fact that this coupling is magnified by a large electric field.

In Section~\ref{sect:comp} we explored the reasons for the observed fast convergence of the resummed moment expansion. We found that, differently from the standard method of moments, the Gaussian factor in the integrand of the definition of the resummed moments suppresses most couplings to higher-order moments, and thereby reduces sensitivity to the approximations made at the truncation level for the non-dynamical higher-order moments. It should be noted, however, that this argument fully holds only in a Bjorken expansion. A weaker argument is expected to apply also for the general (3+1)-dimensional case, but this needs to be validated by future numerical simulations.  

\acknowledgements

This work was supported in part by the U.S. Department of Energy (DOE), Office of Science, Office for Nuclear Physics under Award No. \rm{DE-SC0004286} and through the Beam Energy Scan Theory (BEST) Collaboration, as well as by the National Science Foundation (NSF) within the framework of the JETSCAPE Collaboration under Award No.~ACI-1550223. GV also received support from the Fonds de Recherche du Qu\'ebec -- Nature et Technologies (FRQNT). The work of LT was supported by the Collaborative Research Center CRC-TR 211 ``Strong-interaction matter under extreme conditions'' funded by DFG, and by the Fulbright Program. The work of JN was supported by the Funda\c{c}\~{a}o de Amparo \`a Pesquisa do Estado de S\~ao Paulo (FAPESP) under grants 2016/13517-0 and 2017/05685-2 and also by the Conselho Nacional de Desenvolvimento Cient\'ifico e Tecnol\'ogico (CNPq). UH and JN acknowledge support through a bilateral travel grant from FAPESP and The Ohio State University. UH's research was also in part supported by the ExtreMe Matter Institute EMMI at the GSI Helmholtzzentrum f\"ur Schwerionenforschung, Darmstadt, Germany. Additionally, he would like to thank the Institut f\"ur Theoretische Physik of the J. W. Goethe-Universit\"at, Frankfurt, for their kind hospitality.

\appendix

\section{Four velocity projection and divergence equations}
\label{sect:divergence}

Here we show that the projections onto the four-velocity $u_{\mu_1}$ of the evolution equations (\ref{Red_ev}) for the reducible moments $\Reducible^{\mu_1\cdots\mu_s}_r$ given in (\ref{reducible_moments}) correspond, for generic rank $(r,s)$, to exact equations for the divergence of ${\cal F}_r^{\mu_1\cdots \mu_s}$. Indeed, for $s=\ell{+}1$, Eq.~(\ref{Red_ev}) reads
\begin{eqnarray}
\label{to_project}
  \dot \Reducible^{\alpha\muif{1}{\ell}}_r &+& \deltaReducible_r^{\alpha\muif{1}{\ell}} 
  = r \dot u_\beta \Reducible_{r-1}^{\alpha\beta\muif{1}{\ell}}
\\\nonumber
  &-& \nabla_\beta\Reducible_{r-1}^{\alpha\beta\muif{1}{\ell}} 
  +(r{-}1)\nabla_\rho u_\sigma \Reducible_{r-2}^{\alpha\rho \sigma\muif{1}{\ell}}.
\end{eqnarray}
Since
\begin{equation}
 u_\alpha D = \partial_\alpha -\nabla_\alpha
\end{equation}
one has
\begin{equation}
 u_\alpha\dot \Reducible^{\alpha\muif{1}{\ell}}_r = \partial_\alpha  \Reducible^{\alpha\muif{1}{\ell}}_r - \nabla_\alpha \Reducible^{\alpha\muif{1}{\ell}}_r.
\end{equation}
On the other hand
\begin{equation}
   u_\beta\nabla_\alpha \Reducible^{\alpha\beta\muif{1}{\ell}}_{r-1} 
   = \nabla_\alpha \Reducible^{\alpha\muif{1}{\ell}}_r - \nabla_\alpha u_\beta 
      \Reducible_{r-1}^{\alpha\beta\muif{1}{\ell}}.
\end{equation}
Therefore the projection of Eq.~(\ref{to_project}) onto $u_\alpha$ reads
\begin{eqnarray}
  &&\partial_\alpha  \Reducible^{\alpha\muif{1}{\ell}}_r - \nabla_\alpha  \Reducible^{\alpha\muif{1}{\ell}}_r 
  + \deltaReducible_{r+1}^{\muif{1}{\ell}} 
\nonumber\\
  &&\hspace*{14mm}
  = - \nabla_\alpha  \Reducible^{\alpha\muif{1}{\ell}}_r 
  + r \partial_\alpha u_\beta  \Reducible^{\alpha\beta\muif{1}{\ell}}_{r-1},\qquad
\end{eqnarray}
which simplifies to
\begin{equation}
  \partial_\alpha \dot \Reducible^{\alpha\muif{1}{\ell}}_r  =   r \partial_\alpha u_\beta  \Reducible^{\alpha\beta\muif{1}{\ell}}_{r-1} -\deltaReducible_{r+1}^{\muif{1}{\ell}},
\end{equation}
i.e. an exact equation for the divergence of $\Reducible^{\muif 1\ell}_r$.
In particular, for $\ell{\,=\,}1$ and $r{\,=\,}0$ the contraction of Eq.~(\ref{Tmunu_ev}) with the four velocity corresponds to the conservation of energy and momentum:
\begin{equation}
 \partial_\mu T^{\mu \nu} = -\int dP \, p^\nu\, {\cal C}[f] = 0.
\end{equation}
The same arguments can be used to extend the above result to the Boltzmann-Vlasov equation and to the case of the resummed moments $\Gen^{\muif{1}{s}}_r$ defined in~(\ref{resummed_Gen}).

\section{Derivation of the evolution equations for spatially projected tensor moments}
\label{sect:reduction}

We here derive the exact evolution equations for the spatial tensors $\reducible^{\muif{1}{s}}_r$ from those for $\Reducible^{\muif{1}{s}}_r$. From the definition $\Delta^{\mu\nu} = g^{\mu\nu}{-}u^\mu u^\nu$ it follows that
\begin{equation}
\label{A8}
 \partial_\rho \Delta^\mu_\nu =-u_\nu \partial_\rho u^\mu -u^\mu \partial_\rho u_\nu.
\end{equation}
Some straightforward algebra using the projection property $\Delta^\mu_\nu= \Delta^\mu_\alpha \Delta^\alpha_\nu$ and commuting derivatives with projectors yields
\begin{eqnarray}
\label{A9}
   && \!\!\!\!
   \Delta^{\mu_1}_{\nu_1} \cdots \Delta^{\mu_s}_{\nu_s} \dot \Reducible^{\nu_1\cdots\nu_s}_r =
   \Delta^{\mu_1}_{\nu_1}\Delta^{\nu_1}_{\alpha_1} \cdots \Delta^{\mu_s}_{\nu_s}\Delta^{\nu_s}_{\alpha_s} 
         \dot{\Reducible}^{\alpha_1\cdots\alpha_s}_r 
\nonumber\\
   && \!\!\!\!
    \quad = \dot \reducible^{\langle\mu_1\rangle \cdots \langle\mu_s\rangle}_r 
        + s \, \dot u^{(\mu_1}\reducible_{r+1}^{\muif{2}{s})}, 
\\\nonumber     
\label{A10}
   && \!\!\!\!
   \Delta^{\mu_1}_{\nu_1} \cdots \Delta^{\mu_s}_{\nu_s} \nabla_\alpha \Reducible^{\alpha\nu_1\cdots\nu_s}_r 
\nonumber\\
   && \!\!\!\!
   \quad = \Delta^{\mu_1}_{\nu_1}\Delta^{\nu_1}_{\alpha_1} \cdots \Delta^{\mu_s}_{\nu_s}\Delta^{\nu_s}_{\alpha_s}    
         \nabla_\alpha \Reducible^{\alpha\alpha_1\cdots\alpha_s}_r 
   \\\nonumber 
   && \!\!\!\!
   \quad = \nabla_\alpha \reducible^{\alpha\langle\mu_1\rangle \cdots \langle\mu_s\rangle}_r +\theta 
         \, \reducible^{\muif{1}{s}}_r + s \, \nabla_\alpha u^{(\mu_1}\reducible_{r}^{\muif{2}{s})\alpha},\quad
\end{eqnarray}
where $\theta=\partial{\,\cdot\,}u$ is the scalar expansion rate. Eq.\,(\ref{red_ev}) is then obtained by projecting all free indices with $\Delta$ in Eq.~(\ref{Red_ev}), and moving all terms proportional to the acceleration $\dot u^\mu$ to the right.

\section{On- and off-shell formulations of the Boltzmann-Vlasov equation}
\label{sect:on-of}

Equation~(\ref{RBVE}) is the general Boltzmann-Vlasov equation for phase-space distributions with off-shell momenta. Writing explicitly the momentum dependence of the distribution function, it reads:
\begin{equation}
\begin{split}
  & p\cdot\partial f(x,p) + m(x) \bigl(\partial_\rho m(x)\bigr) \partial^\rho_p f(x,p)  
\\
  & \ \ \qquad \qquad + q F_{\alpha\beta}p^\beta\partial^\alpha_p f(x,p) =-{\cal C}[f](x,p),
\label{RBVE_explicit}
\end{split}
\end{equation}
where $p_\mu$ is a generic momentum four-vector. While classical kinetic theory considers only on-shell particles, the  form (\ref{RBVE_explicit}) is attractive because of its manifest covariance, with a distribution function that transforms as a scalar under Lorentz transformations.

The on-shell part of the distribution function is
\begin{equation}
\label{f_on}
 \fon(x, \hat p) = \int dp_0 \, 2 \,\Theta (p_0)\, \delta(p^2{-}m^2)\, f (x,p),
\end{equation}
where $\hat p = (\hat  p_0,{\bf p})$ the on-shell four-momentum vector with 
\begin{equation}
\label{p0_on}
   \hat p_0(x,{\bf p})  = -\frac{1}{g^{00}}\left[ g^{0i}p_i 
   - \sqrt{\left( g^{0i}p_i \right)^2{-}g^{00}\left( g^{ij}p_i p_j{-}m^2 \right)} \right]\!.
\end{equation}
The $x$ dependence of $\hat p_0$ comes from the metric tensor which in general is not constant. To derive from (\ref{RBVE_explicit}) the on-shell Boltzmann-Vlasov equation, it is convenient to write the equation in generic (non-Cartesian) coordinates. Particle momenta are covariant vectors $p_\mu$, therefore they transform with the inverse of the Jacobian matrix, $p'_\mu=({\cal J}^{-1})^\alpha_\mu p_\alpha$, where primes indicate the transformed coordinates and momenta. In general the Jacobian itself is $x$-dependent, so the components of the momenta in the new frame also depend on position, $p'_\mu(x')$. The Vlasov term in Eq.~(\ref{RBVE_explicit}) does not change, since the indices are dummy indices that are being summed over:
\begin{equation}
  m \partial_\rho m \partial^\rho_p f+ q F_{\alpha\beta}p^\beta\partial^\alpha_p f 
  \longrightarrow  m \partial'_\rho m \partial^\rho_{p'}  f + q F_{\alpha\beta}p^{\prime\beta}
  \partial^\alpha_{p'} f. 
\nonumber \\
\end{equation}
The derivative with respect to the coordinates is different. Since in the transformed frame the (Lorentz scalar) distribution function $f\bigl(x',p'(x')\bigr)$ depends on $x'$ through both arguments, the gradient in the momentum direction now reads
\begin{eqnarray}
\label{eq:C5}
 && p^{\prime\mu}\partial'_\mu f\bigl(x', p'(x')\bigr)  
\\\nonumber
 &&= p^{\prime\mu} \Bigl[\partial'_\mu f\bigl(x', p'(x')\bigr)\big|_{p'} 
                                         +\Bigl(\frac{\partial p'_\nu}{\partial x^{\prime\mu}}\Bigr) \partial_{p'}^\nu f(x',p')\big|_{x'}
                                 \Bigr]. 
\end{eqnarray}
The partial derivatives of the momenta can be rewritten using the tensor equation (valid in any reference frame)
\begin{equation}
  d_\mu p_\nu = \partial_\mu p_\nu -\Gamma^\alpha_{\mu\nu}p_\alpha =0 \ 
  \Longrightarrow\  \partial_\mu p_\nu = \Gamma^\alpha_{\mu\nu}p_\alpha
\end{equation}
where $d_\mu$ denotes the covariant derivative. Summing everything and dropping the primes one finds
\begin{eqnarray}
\label{RBVE_generic}
 &&p\cdot\partial f + \Gamma^\alpha_{\nu\mu} \, p_\alpha p^\mu\,  \partial^\nu_p  f  
 + m (\partial_\rho m) \partial^\rho_p  f 
 + q F_{\alpha\beta}p^\beta\partial^\alpha_p f 
\nonumber\\
 &&\qquad = -{\cal C}[f].
\end{eqnarray}
Equation~(\ref{RBVE_generic}) is well defined even in non-flat space-time. However, in the absence of a Cartesian frame (i.e. a global map to Minkowski space) one needs a more refined mathematical framework to obtain it. We just mention the classical kinetic theory based approach presented in \cite{2009PhyA..388.1818D} which can also be extended to the Boltzmann-Vlasov case. Alternatively, the relativistic Boltzmann equation can be extracted from an effective quantum field theory as the classical (non-quantum) limit of the evolution of the Wigner functions \cite{degroot}, which reproduces the same result \cite{Fonarev:1993ht}.

Using the definitions~(\ref{f_on}) and~(\ref{p0_on}), compatibility of the Levi-Civita connection with the metric,
\begin{equation}\begin{split}
 0=d_\alpha g_{\mu\nu} & = \partial_\alpha g_{\mu\nu} -\Gamma^\beta_{\alpha\mu}g_{\beta\nu}-\Gamma^\beta_{\alpha\nu}g_{\beta\mu}  \\
 & \qquad \Longrightarrow\ \partial_\alpha g_{\mu\nu} = \Gamma^\beta_{\alpha\mu}g_{\beta\nu} + \Gamma^\beta_{\alpha\nu}g_{\beta\mu},\end{split}
\end{equation}
and the delta distribution identity
\begin{equation}
 2p^\mu\delta^\prime(p^2{-}m^2) = -\partial^\mu_p \delta(p^2{-}m^2),
\end{equation}
one finds after tedious but straightforward algebra that the $p_0$ integral of the off-shell equation~(\ref{RBVE_generic})
%
\begin{eqnarray}
\!\!\!\!\!\!
&&\!\!\!\!\!\!
 \int d p_0 \, 2 \Theta(p_0) \delta(p^2{-}m^2) \, p^0 \Bigl[  p\cdot\partial f + \Gamma^\alpha_{\nu\mu} \, 
 p_\alpha p^\mu\,  \partial^\nu_p  f  
\nonumber\\
&&\hspace*{35mm}
+\, m (\partial_\rho m) \partial^\rho_p  f+ q F_{\alpha\beta}p^\beta\partial^\alpha_p f \Bigr]
\nonumber\\
&&\!\!\!\!\!\!
  = - \int dp_0 \, 2 \Theta(p_0) \delta(p^2{-}m^2) \, p^0\, {\cal C}[f]
 \equiv - {\cal C}_{\rm on.}[\fon]
\end{eqnarray}
%
corresponds exactly to
\begin{equation}
  \hat p^\mu \partial_\mu \fon + \Gamma^\alpha_{\mu i}\, \hat  p^\mu \hat p_\alpha \,  \partial^i_p  \fon  
 + q \, F_{i\nu} \, \hat p^\nu \, \partial^i_p  \fon  =  -{\cal C}_{\rm on}.
\end{equation}
Indeed, Eq.~(\ref{RBVE_on}) is exactly the last equation, but in a Cartesian reference frame and with the understanding that all four-momenta are on-shell being implicit, i.e. not denoted by hats.

A different but physically equivalent convention is to consider the contravariant momenta $p^\mu$ as the fundamental ones, i.e.  assuming a vanishing partial derivative for the contravariant components $p^\mu$ but not for the covariant ones, $p_\mu=g^{\mu\nu} (x) p_\nu$. In this case the coupling to the connection coefficients reads
\begin{equation}
 -\Gamma^{\alpha}_{\mu\nu}p^\mu p^\nu \partial_\alpha^p f
\end{equation}
for the off-shell case and
\begin{equation}
 -\Gamma^{i}_{\mu\nu}\hat p^\mu \hat p^\nu \partial_i^p \fon
\end{equation}
for the on-shell case. In spite of the two conventions being equivalent we find it more convenient here to use Eq.~(\ref{RBVE_on}) because in the Milne coordinate system the contribution from the connection coefficients simplifies exactly when adopting the covariant moments convention, while it does not for the contravariant one.

\section{Treatment of the negative energy index moments}
\label{sect:irreducible}

Based on completeness arguments involving irreducible tensors \cite{Denicol:2012cn} it is possible to approximate the $\reducible_r^{\muif 1s}$ moments having a negative energy index $r$ with a non-dynamical series of moments of non-negative $r$. This has been a very successful approach when deriving second order viscous hydrodynamics from the Boltzmann equation, in which case one can restrict one's attention to rank-2 tensors only \cite{Denicol:2012cn}. This approach can, however, quickly become numerically too expensive in more general situations. In this section we check numerically the convergence and stability of the expansion in a simple case. 

\begin{figure}[t]
\centering
\includegraphics[angle=0,width= \columnwidth]{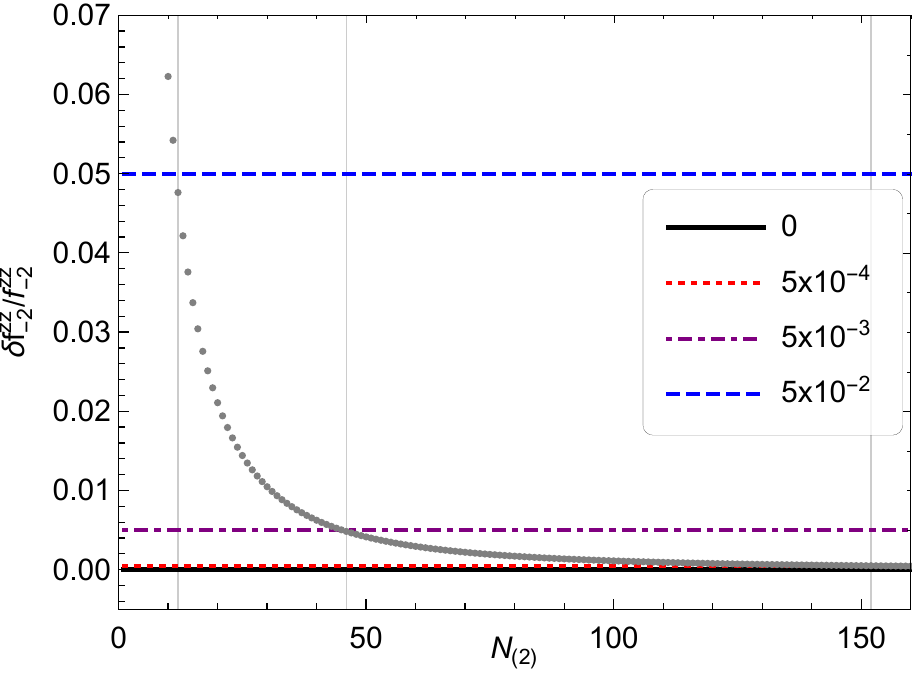}
\caption{
	(Color online ) Convergence of the series in~(\ref{sum}) for different truncations $N_{(2)}$. 
	The vertical grey lines correspond to the minimum $N_{(2)}$ after which all the points are 
	below the horizontal lines. }
\label{F10}
\end{figure}

To illustrate the problem let us look at Eq.~(\ref{red_ev_BV}), assuming a system of massless particles. At leading order one encounters the $\reducible_{-2}^{\mu\nu}$ moment e.g. in the evolution equation for the particle diffusion (or electric) current (i.e. for the $\reducible$-moment with $r=0$ and $s=1$). We consider, as an example, an anisotropic distribution function of the form
\begin{equation}
 f = e^{(\mu-p\cdot u)/T} \left( 1 - P_2\left( \frac{p\cdot z}{p\cdot u} \right)\right),
\end{equation}
with $P_2$ being the second Legendre polynomial \cite{NIST:DLMF}. Because of the matching conditions, the particle and energy densities are the same as for the equilibrium distribution with chemical potential $\mu$ and temperature $T$. The same argument holds for all the $\reducible_n$ and $\reducible_n^\mu$ moments. The higher ranking tensor moments, however, differ from their equilibrium expectation values. In particular,
\begin{equation}
  \reducible_{-2}^{\mu\nu}z_\mu z_\nu \equiv \reducible_{-2}^{zz}= \frac{1}{5}\reducible_0\big|_{\rm eq.} 
  \neq \frac{1}{3}\reducible_0\big|_{\rm eq.}=\reducible_{-2}^{zz}\big|_{\rm eq.}.
\end{equation}
The prescription given by the authors of Ref.~\cite{Denicol:2012cn} is probably the best non-dynamical one that can be used to approximate the moments with negative energy index, being based on a well-tested polynomial expansion of the factor $(p\cdot u)^{-|r|}$ in the definition of the moments that makes use of the orthogonality relations of the irreducible basis. Following their prescription, in this particular case the moment $\reducible_{-2}^{zz}$ can be approximated by the series
\begin{eqnarray}
\!\!\!\!
  &&\reducible_{-2}^{zz} = \frac{1}{5} \reducible _0 \big|_{\rm eq.} 
\\\nonumber
  &&\quad \simeq  \reducible _0 \big|_{\rm eq.} 
  \left[ \frac{1}{3} -\frac{12}{15}\sum_{n=0}^{N_{(2)}}(-1)^n\frac{ (n + 3)!}{(n + 5)! n!} 
          \sum_{m=n}^{N_{(2)}} \frac{(m+1)!}{(m-n)!} \right]\!,
\end{eqnarray}
with $N_{(2)}$ being the maximum energy index of irreducible tensors of rank two considered, as explained in \cite{Denicol:2012cn}. In this particular case the relative error therefore reads
\begin{equation}
\label{sum}
\begin{split}
 &\frac{\reducible_{-2}^{zz}\big|_{\rm approx}}{\reducible_{-2}^{zz}\big|_{\rm exact}} =  \left[ \frac{5}{3} -4\sum_{n=0}^{N_{(2)}}(-1)^n\frac{ (n{+}3)!}{n!\,(n{+}5)! } \sum_{k=n}^{N_{(2)}} \frac{(k{+}1)!}{(k{-}n)!} \right], \\ \\
 &\Rightarrow \frac{\delta f_{-2}^{zz}}{f_{-2}^{zz}} = \frac{\reducible_{-2}^{zz}\big|_{\rm approx}}{\reducible_{-2}^{zz}\big|_{\rm exact}} -1 =\\
  &  = \left[ \frac{2}{3} -4\sum_{n=0}^{N_{(2)}}(-1)^n\frac{ (n{+}3)!}{n!\,(n{+}5)! } \sum_{k=n}^{N_{(2)}} \frac{(k{+}1)!}{(k{-}n)!} \right].
 \end{split}
\end{equation}
In Fig.~\ref{F10} we show the relative precision of the approximation for different truncation orders $N_{(2)}$. It is necessary to consider $12$ irreducible moments ($N_{(2)}=12$) of positive energy index to reproduce $\reducible_{-2}^{zz}$ with $10^{-2}$ accuracy (relative error smaller than $5\%$). One needs $N_{(2)}=46$ to reach $10^{-3}$ precision, and $N_{(2)}=152$ for $10^{-4}$.

In the particular case of interaction with an electromagnetic field, the term proportional to $\reducible^{zz}_{-2}$ can easily become the dominant one in the evolution of the electric current, and this indeed happens in the cases we tested numerically in this paper. If the dominant contribution on the r.h.s. of Eq.~(\ref{red_ev_BV}) is not reproduced at the desired precision, it is very unlikely that the desired precision can still be achieved for dynamically evolved moment on the l.h.s. 

This example shows that the treatment of the moments with negative energy index proposed in \cite{Denicol:2012cn} can become numerically costly, because of the large number of degrees of freedom that must be taken into 
account.\footnote{%
	In \cite{Denicol:2016bjh}, on the other hand, a precision of the order of $10^{-3}$ in the final solutions 
	is obtained already with a handful of degrees of freedom, by using a modified set of moments. 
	}
Moments with higher tensor rank have the same problems. While this polynomial series has been very helpful for the lowest order of the expansion needed to obtain Israel-Stewart-like viscous hydrodynamic theories, and for situations very close to local thermal equilibrium, it is not convenient for the analysis done in this paper.

As a final remark we note that making use of the resummed moments presented in Sect.~\ref{sect:Res} corresponds to having no truncation in Eq.~(\ref{sum}) at all, in the sense of taking $N_{(l)}\to\infty$. The numerical precision of the code, however, limits the accessible information on the $\reducible_r^{\muif 1s}$ moments for very large $r$ values since those are represented by many consecutive derivatives of the resummed $\gen^{\muif 1s}_1$ moments.
%
%

\section{Exact solutions of the coupled Boltzmann-Vlasov-Maxwell equations}
\label{sect:exact_BVM_equations}

In this appendix we will consider the equation for the on-shell distribution functions of a multi-particle gas in Milne coordinates. We will show that it is necessary to have at least a two component system (particles and anti-particles) in order to have a longitudinally boost invariant and transverse homogeneous expansion and, at the same time, fulfill the Maxwell equations.

According to Eq.~(\ref{RBVE_on}), making use of the simplifications between the connection coefficients terms, one finds in Milne coordinates (just like in the Cartesian ones)
\begin{equation}
 p\cdot\partial f + q F_{i\alpha}p^\alpha \frac{\partial f}{\partial p_i} = -{\cal C}.
\end{equation}
We omitted the ``on" in $f_{\rm on}$ and the hat in $\hat p$ because in this section we will consider only the on-shell version of the Boltzmann-Vlasov equation.

If one assumes that particles have a longitudinally boost invariant and transversely homogeneous distribution one has $f = f(\tau, p_T, p_\eta)$. In particular, in order to have the so called Bjorken expansion, one must add the requirement $f(\tau, p_T, p_\eta)=f(\tau, p_T,- p_\eta)$, i.e. $Z_2$ symmetry. Therefore,
\begin{equation}
 \partial_\mu f(\tau, p_T, p_\eta) = \delta^\tau_\mu \, \partial_\tau f(\tau, p_T, p_\eta),
\end{equation}
\begin{equation}
 \frac{\partial f}{\partial p_x} =\frac{\partial (p_T =\sqrt{p_x^ +p_y^2})}{\partial p_x} \frac{\partial f}{\partial p_T} = \frac{ p_x}{p_T} \frac{\partial f}{\partial p_T},
\end{equation}
\begin{equation}
 \frac{\partial f}{\partial p_y} =\frac{\partial (p_T =\sqrt{p_x^ +p_y^2})}{\partial p_y} \frac{\partial f}{\partial p_T} = \frac{ p_y}{p_T} \frac{\partial f}{\partial p_T}.
\end{equation}
The Vlasov term couples the fields to the particles and only some fields configurations do not spoil the assumed symmetry of the distribution function. Indeed, using the last expression and decomposing the tensor $F_{\mu\nu}$ along the time direction in Milne coordinates (the Bjorken four-velocity) one finds
\begin{eqnarray}
   &&\!\!\!\!\!\!
   qF_{i\alpha}p^\alpha\partial^i_p  f(\tau,p_T,p_\eta) = q p^\tau \! \left[ \frac{E_x p_y{+}E_y p_x}{p_T} 
   \frac{\partial  f}{\partial p_T} + E_\eta \frac{\partial f}{\partial p_\eta} \right] 
\nonumber\\ 
  &&
   + q \left[  p^\eta \frac{B^x p^y{-}B^y p^x}{p_T} \frac{\partial  f}{\partial p_T} 
   + \bigl( p^x B^y{-}p^y B^x\bigr) \frac{\partial  f}{\partial p_\eta} \right].
\end{eqnarray}
The electromagnetic tensor $F_{\mu\nu}$ and, therefore, the electric and magnetic fields do not depend on the momenta of particles. The components of the electric $E^\mu$ and magnetic $B^\mu$ fields are just space-time functions and, therefore, they cannot compensate any angular dependence in the momentum space caused by the explicit terms in $p_x$ and $p_y$ instead of the invariant $p_T = \sqrt{p_x^2 + p_y^2}$. In fact, transverse fields break homogeneity in the transverse plane explicitly and are, thus, not allowed. Longitudinal magnetic fields do not couple at all with the particles in this case and, therefore, they do not break any symmetry.

If the distribution function has to be even in the longitudinal momentum $p_\eta$ ($Z_2$ symmetry), no electric fields are allowed since $\partial f/\partial p_\eta$ must be odd and the electric field is momentum-independent and cannot regularize this term. If one relaxes the traditional Bjorken symmetry, maintaining the longitudinal boost invariance and the transverse homogeneity, but relaxing the $Z_2$ requirement, an electric field is allowed. However, only a longitudinal component is allowed, i.e. $E_\mu = \delta_\mu^\eta E_\eta(\tau)$.

This simple analysis of the Vlasov term shows that the the only form of $F_{\mu\nu}$ which does not explicitly break the (relaxed) symmetry requirements is
\begin{equation}
 F_{\mu\nu} = \left( \vphantom{\frac{}{}} \delta^\eta_\mu \delta^\tau_\nu - \delta^\eta_\nu \delta^\tau_\mu \right) E_\eta(\tau) + \varepsilon_{\mu\nu\tau \eta} B^\eta(\tau, x, y, \eta).
\end{equation}
If the electromagnetic field is not external, and there is no other source than the particles in the plasma, the Maxwell equations provide some additional constraints. Indeed, the Bianchi identities read
\begin{eqnarray}
  &&\!\!\!\!\!
   \varepsilon^{\mu\nu\rho\sigma}\partial_\nu F_{\rho\sigma} 
   = \varepsilon^{\mu\nu\rho\sigma}\left( 2\, \delta^\eta_\rho \, \delta^\tau_\sigma \, \partial_\nu E_\eta 
     +\varepsilon_{\rho\sigma\tau\eta} \partial_\nu B^\eta \right) 
\nonumber\\
  && = 2\left( \delta^\mu_\tau\delta^\nu_\eta -\delta^\mu_\eta \delta^\nu_\tau \right)\partial_\nu B^\eta
       = 2\delta^\mu_\tau \partial_\eta B^\eta -2 \delta^\mu_\eta \partial_\tau B^\eta.\qquad
\end{eqnarray}
The two non-trivial equations (the longitudinal and the time projections) require that $B^\eta = B^\eta(x,y)$.

The coupling to particles is less trivial but it can be solved with some algebraic manipulations
\begin{eqnarray}
  &&\!\!\!\!\!\!
  J^\nu = d_\mu  F^{\mu\nu} = d_\mu\left[ E^\mu u^\nu -E^\nu u^\mu 
            + \varepsilon^{\mu\nu\rho\sigma} u_\rho B_\sigma \right]  
\nonumber\\\nonumber
 &&\!\!\!\!\!\!
  = (d\cdot E) u^\nu +E^\mu d_\mu u^\nu - u^\nu d_\mu E^\nu -\theta E^\nu
\\\nonumber
 &&  +\, \varepsilon^{\mu\nu\rho\sigma}\partial_\mu( u_\rho B_\sigma) 
       +\Gamma^\mu_{\mu\alpha}\varepsilon^{\alpha\nu\rho\sigma}u_\rho B_\sigma 
       +\Gamma^\nu_{\mu\alpha}\varepsilon^{\mu\alpha\rho\sigma}u_\rho B_\sigma 
\\ \nonumber 
 &&\!\!\!\!\!\!
     = E^\eta \Gamma^\nu_{\eta \tau} -\delta^\nu_\eta \partial_\tau E^\eta 
        - \Gamma^\nu_{\tau\eta} E^\eta -\theta \delta^\nu_\eta E^\eta 
       + \delta^\nu_y \partial_x B_\eta
\\\nonumber
   && -\,\delta^\nu_x \partial_y B_\eta  
       + \delta^\nu_y \Gamma^\mu_{\mu x} B_\eta 
       - \delta^\nu_x \Gamma^\mu_{\mu y} B_\eta 
\\
 &&\!\!\!\!\!\!
   = -\delta^\nu_\eta \left( \partial_\tau E^\eta +\theta E^\eta \right) +\delta^\nu_y  \partial_x B_\eta -\delta^\nu_x \partial_y B_\eta.
\end{eqnarray}
Using the fact that $E^\eta =g^{\eta\eta} E_\eta = -1/\tau^2 E_\eta$, the last four independent equations read
\begin{equation}
 J^\tau \equiv J_\tau = 0, \ 
 \partial_y B^\eta =0, \ 
 \partial_x B^\eta =0, \ 
 \partial_\tau E_\eta = \frac{1}{\tau}E_\eta{-}J_\eta.
\end{equation}
The magnetic field $B^\eta$ must be longitudinal and constant, while the electric field feels the backreaction from the particles. The time component of the electric current must be zero and, therefore, the electric current is space-like and cannot be used to define the Eckart frame. In particular, it is not possible to have a single charge carrier species. The simplest case is to consider a gas of particles (with distribution $f$) and anti-particles (with distribution $\bar f$) following the equations of motion
\begin{equation}
 \partial_\tau f + q E_\eta \frac{\partial f}{\partial p_\eta} = -\frac{1}{p^\tau} {\cal C}[f,\bar f],
\end{equation}
\begin{equation}
 \partial_\tau \bar f - q E_\eta \frac{\partial \bar f}{\partial p_\eta} = -\frac{1}{p^\tau}\bar {\cal C}[f, \bar f],
\end{equation}
\begin{equation}
 \partial_\tau E_\eta = \frac{1}{\tau} E_\eta - q \tau \int \frac{ d^3 p}{\tau} \, \frac{p_\eta}{\tau p^\tau} \left( \vphantom{\frac{}{}} f{-}\bar f \right), 
\end{equation}
\begin{equation}
 \int \frac{ d^3 p}{\tau}  \left( \vphantom{\frac{}{}} f{-}\bar f \right)=0 .
\end{equation}
In order to solve such system of equations it is necessary to specify the collisional kernel. The simplest situation is the relaxation time approximation used in this work
\begin{eqnarray}\nonumber
  &&{\cal C} = \frac{(p\cdot u)}{\taueq}\bigl(f{-}\feq\bigr), \qquad  
 \bar{\cal C} = \frac{(p\cdot u)}{\taueq}\bigl({\bar f}{-}{\bar f}_{\rm eq} \bigr), \\ 
\nonumber
 &&  \feq =  \exp\left[ \frac{\mu-p\cdot u}{T} \right], \quad
       {\bar f}_{\rm eq} =\exp\left[ \frac{-\mu-p\cdot u}{T} \right].
\end{eqnarray}
The four velocity must be the time-like eigenvector of the stress-energy tensor as defined in the Landau frame, otherwise the collision kernel does not conserve energy and momentum. The effective chemical potential $\mu$ and temperature $T$ are defined through the matching conditions
\begin{eqnarray}
 && \int_p \, (p\cdot u)^2 (f{+}\bar f) \equiv  \int_p \, (p\cdot u)^2 (f_{\rm eq}{+}\bar f_{\rm eq}) 
       = \ped_{\rm eq}(\mu,T),  \nonumber\\ \\\nonumber
 &&q \int_p \, (p\cdot u) (f{-}\bar f) \equiv  q\int_p \, (p\cdot u) (f_{\rm eq}{-}\bar f_{\rm eq}) = \rho_{\rm eq}(\mu,T).
\\
\end{eqnarray}
In general, because we dropped the $Z_2$ symmetry, the four-velocity $u^\mu$ is not necessarily the time direction in Milne coordinates as it occurs in the full Bjorken symmetry group \cite{Gubser:2010ze}. The time direction is not any longer the only vector that fulfills the (relaxed) symmetry requirements. However, one can assume for simplicity that  it exists a solution in which $u^\mu=(1,0,0,0)$ and verify that, using an initial distribution that fulfills this requirement (for instance, local equilibrium), the four velocity remains the same throughout the evolution. Under this assumption the chemical potential must be vanishing, since $J_\tau = J\cdot u$. The remaining equations are indeed the ones in~(\ref{particle_BV}), (\ref{antiparticle_BV}) and~(\ref{exact_E_Maxwell}). The solution (as one can easily verify) is the one in Eqs.~(\ref{part}) and~(\ref{antipart}).

The Landau definition of the four-velocity can be rewritten as follows:
\begin{equation}
 0= \int \frac{d^3 p}{\tau}p_i\, \bigl(f{+}\bar f \bigr).
\end{equation}
The initial conditions $f_0$ and $\bar f_0$ fulfill the requirement by hypothesis. The contribution for the momentum shift simplifies exactly when considering the particle and antiparticle terms, as can be seen by splitting the integral in a particle and an antiparticle contribution and performing a change of variables $p_\eta \pm \Delta p_\eta\to p_\eta$. After a similar manipulation of the remaining contribution, one can use the fact that the local equilibrium distribution is invariant under parity and prove the general statement. The very same arguments can be used to show that, if $J_\tau = u\cdot J$ vanishes at the initial condition, it must vanish at all times.


\bibliography{EM_Boltz}

\end{document}